%% file: Zeta.tex
\RequirePackage{ifpdf}
\documentclass[hyper,12pt,letterpaper]{JHEP3}
\pdfoutput=1
\usepackage{graphicx}
\usepackage{amsmath,amssymb,array}

\input{gdefn.tex}

\numberwithin{equation}{section}

\usepackage{float}
\usepackage[font=small,labelfont=bf]{caption}
\usepackage{mathrsfs}
\usepackage[parfill]{parskip}
\usepackage{subcaption}

\title{Phase Space Distribution of Riemann Zeros}
\author{Parikshit Dutta\footnote{parikshitdutta@yahoo.co.in} $^{1,2}$\ and Suvankar
  Dutta\footnote{suvankar@iiserb.ac.in} $^{1}$\\
  $^{1}$Department of Physics \\
   Indian Institute of Science Education and Research Bhopal \\
   Bhopal 462 066, 
   India\\
$^{2}$ School of Physical Sciences\\
  Jawaharlal Nehru University\\
  New Delhi 110067, India}

\abstract{We present the partition function of a most generic $U(N)$
  single plaquette model in terms of representations of unitary
  group. Extremising the partition function in large N limit we obtain
  a relation between eigenvalues of unitary matrices and number of
  boxes in the most dominant Young tableaux distribution. Since,
  eigenvalues of unitary matrices behave like coordinates of free
  fermions whereas, number of boxes in a row is like conjugate momenta
  of the same, a relation between them allows us to provide a phase
  space distribution for different phases of the unitary model under
  consideration.  This proves a universal feature that all the phases
  of a generic unitary matrix model can be described in terms of
  topology of free fermi phase space distribution. Finally, using this
  result and analytic properties of resolvent that satisfy
  Dyson-Schwinger equation, we present a phase space distribution of
  unfolded zeros of Riemann zeta function.}

\begin{document}

\maketitle

\section{Introduction and summary}

Hilbert and P\'oyla speculated that there is spectral origin for
non-trivial (also called unfolded) zeros of \rzf.  This means, if
Riemann hypothesis is true {\it i.e.} non-trivial zeros of \rzf \
$\zeta(s)$ lie on $\Re [s]=1/2$ line, (which mathematically means
$\zeta(1/2+i \ t_i)=0$ for $t_i\in \mathbb{R}$), then $t_i$
corresponds to eigenvalues of an unbounded self-adjoint operator.

In 1972, H. Montgomery conjectured that pair correlation function
between unfolded zeroes of zeta function to be same as the pair
correlation function of random matrices taken from a Gaussian Unitary
Ensemble \cite{Montgomery}.  In fact there exists a strong similarity
between eigenvalues and unfolded zeros. Namely, they show repulsion at
short range.  Unfolded zeros repel each other quadratically which is
similar to repulsion between eigen values of a random matrix, coming
from the Harr measure.  Montgomery's conjecture led to numerous study
on the analogy between the distribution of the zeroes of zeta function
and random matrix theory. Berry and Keating showed that there exists a
similarity between the formula for counting function of heights
$t_{i}$ of unfolded zeroes of zeta function and energy eigenvalues
$\tilde{E}_{n}$ of some unknown hermitian operator (in spirit of the
Hilbert Polya conjecture) \cite{Berry-Keating}.  In \cite{Keating}
J. Keating\footnote{See also \cite{Keating2}.} compare the value
distribution of the logarithm of the zeta function on the critical
line to the expectation value of the characteristic polynomial of a
random unitary matrix under the Harr measure. Thus the similarity
between the the eigen values of a unitary matrix (or Hermitian matrix)
and the non trivial zeroes of the zeta function has long been known
and studied.

One of the goals of this paper is to construct a phase space
distribution for unfolded zeros of \rzf. It is well known that
eigenvalues of unitary matrices behave like positions of free fermions
whereas representations of $U(N)$, at the same time, have an
interpretation in the language of free fermions with number of boxes
of Young tableaux being like momentum \cite{douglas}.  Thus, a
relation between eigenvalues and number of boxes of a Young diagram
provides a phase space picture of different phases of a unitary matrix
model. Different phases are distinguished by different topologies of
droplets \cite{Dutta:2007ws, dutta-dutta}. In this paper we first
attempt to construct a unitary matrix model whose eigenvalue
distribution for a particular phase captures information of
distribution of unfolded zeros of \rzf. Then, we find the dominant
representation(s) ({\it i.e.} number of boxes in different rows of a
diagram) corresponding to that particular phase of the model. We
observe that there exists a natural relation between eigenvalues and
number of boxes. This enables us to draw a phase space distribution of
unfolded zeros of \rzf.

Logical flow of our paper goes as follows. We start with the following
partition function for one plaquette model
\be\label{eq:plaq}
{\cal Z} = \int [{\cal D}U] \exp\lB \sum_{n=1}^{\infty}
{\b_n\over n}\lb\Tr \ U^n +\Tr\ U^{\dagger n}\rb\rB,
\ee
where $U$ is a $N\times N$ unitary matrix and $[\cD U]$ is
corresponding Haar measure. The model shows a particular distribution
$\rho(\{\theta_{\a}\})$ of its eigenvalues depending on various
parameters $\b_n$ of the theory.

{\it The first goal of this paper} is to write down a unitary one
plaquette model such that eigenvalue distribution function captures
information of unfolded zeros of \rzf. In particular, we find a no-gap
solution (eigenvalues are distributed between $0$ and $2\pi$) where
$\r(\q)$ has logarithmic divergences (spike like solution) at the
positions of unfolded zeros of \rzf. Note that we are not mapping the
eigenvalue distribution with the distribution of unfolded zeros,
rather we are mapping divergences of eigenvalue distribution with the
zeros of \rzf.

{\it The second goal and the most important observation} of this paper
is following. The integral over unitary matrices in equation
(\ref{eq:plaq}) can as well be done in a different way. Expanding the
exponential in terms of character $\chi$ of conjugacy classes of a
symmetric group and using the orthogonality relation between the
characters of $U (N )$ in different representations, it is possible to
write down the partition function in terms of a sum over
representations of unitary group of rank $N$
\cite{Dutta:2007ws}. Since representations of a unitary group can be
described by Young diagrams, this is, therefore, equivalent to finding
the Young tableaux distribution which dominates the partition function
in the large $N$ limit. It was suggested in \cite{Dutta:2007ws} that
in the large $N$ limit unitary matrix models admit a phase space
description. This follows from the fact that eigenvalues of the
holonomy matrix behave like position of free fermions whereas, the
number of boxes in the Young tableaux are like conjugate
momentum. Which means eigenvalue distribution is like position
distribution and arrangement of boxes in the Young diagram is like
momentum distribution for $N$ non-interacting
fermions. \cite{Dutta:2007ws} found an identification between these
two descriptions which essentially led to free fermion picture.

This free fermi picture is very intriguing. One can ask why there
should be a free fermion picture for large N saddle points of an
interacting theory. Does there exist an underlying many particle
non-interacting quantum description for each large $N$ saddle point?
Although we do not have clear understanding of these questions but
before we try to attempt these, a natural question arises at this
point if this free fermi phase space description is universal, {\it
  i.e.}, if all large $N$ saddle points have an underlying free fermi
picture.

The observed relation in \cite{Dutta:2007ws} between eigenvalues and
number of boxes in a Young tableaux was some what accidental. There
was no natural way to arrive this relation. Therefore, universality of
free fermi picture {\it i.e.} existence of a free fermi phase space
distribution for any class of solution, remains under question. In
this paper we explicitly show that relation between $h$ and $\q$
follows from extremization condition in the large $N$ limit. This
proves that a free fermi description is always possible for any class
of solution or phase of a unitary matrix model. Although, we have
proved this for one plaquette model, but it is easy to extend our
proof to include the models considered in
\cite{Jurkiewicz:1982iz}. This is the most important result of this
paper.

Finding out the most dominant Young representation in large $N$ limit
is a technically challenging problem in general. This is because we do
not have an exact formula for character of permutation groups in
presence of arbitrary number of cycles. The most useful formula of
character is given by Frobenius. However, this formula involves $N$
number of auxiliary variables.  In presence of one cycles only, one
can show that it is possible to get rid of these auxiliary variables
and find an expression for character in terms of number of boxes in a
diagram. \cite{Dutta:2007ws} used this expression for character
(because the model they considered in that paper corresponds to
existence of one cycle only) and found the most dominant
representations for different classes of solutions. But for a generic
unitary matrix model like (\ref{eq:plaq}) the character involves
arbitrary number of cycles, then it is technically difficult to
extremize the partition function and obtain a Young tableaux
density. In this paper we show that although it is difficult to find
the most dominant Young tableaux density for a generic unitary matrix
model, but one can definitely find a relation between number of boxes
$h_i$ in a representation and eigenvalues $\q_i$ of a unitary matrix.

Proving the existence of free fermi description for any generic class
of solution of a unitary matrix model, we proceed to our {\it third
  and final goal}, {\it i.e.} to draw a phase space distribution for
unfolded zeros of \rzf. It turns out that the phase space has a shape
of a broom (in two dimensions of course). There are infinite number of
sticks on the right. Density of these sticks increases as one moves
towards $\pi=0$. These sticks corresponds to positions of unfolded
zeros.

It has been observed in \cite{Berry-Keating} that if $t_n$ (hight of
critical line) is analogous to eigenvalue of some Hamiltonian, then
the corresponding conjugate variable is time period of some primitive
orbits. This time period is proportional to $\ln p$ where $p$ is prime
number. In our analysis we see that eigenvalue distribution is related
to distribution of unfolded zeros, therefore we expect that the Young
distribution function captures some information about prime number
distribution. Although our understanding regarding this is not yet
complete, but in this paper we present some clue or hint about this.

Organization of this paper is following.
\begin{itemize}
\item In section \ref{sec:plaq} we discuss the properties of resolvent
  in the context of a most generic single plaquette model.
\item Important properties of \rzf\ and useful relations are discussed
  in section \ref{sec:rzf}
\item In section \ref{sec:model} we present the unitary plaquette
  model in details and show how eigenvalue distribution function
  captures information of unfolded zeros.
\item In section \ref{sec:Young} we prove the universal feature that
  all the phases of a generic unitary matrix model can be described in
  terms of topology of free fermi phase space distribution.
\item Phase space distribution of unfolded zeros has been plotted in
  section \ref{sec:phasespace}. Also the relation between Young
  distribution and prime number distribution has been studied in this
  section.
\item Finally, in the last section (section \ref{sec:dis}) we discuss
  some open problems related to this work.

\end{itemize}

\input{plaq.tex}
\input{rzf.tex}
\input{model.tex}

\input{young.tex}

\input{phspace-prime.tex}

\section{Discussion}
\label{sec:dis}

In this paper we find a universal feature that all the phases of a
generic unitary matrix model can be described in terms of topology of
free fermi phase space distribution. This is therefore, the right time
to get back to the original question which we raised in the
introduction : ``Does there exist an underlying many particle
non-interacting quantum description for each large N saddle point?'' A
series of papers \cite{Dhar:1992rs, Dhar:1992hr, Dhar:1992cs,
  Dhar:1993jc} by Dhar, Mandal and Wadia will be useful to understand
this question in details. It would be interesting to see if one can
formulate a quantum theory from classical (large $N$) phase space
description. The difference between free fermi droplets obtained for
different phases of unitary matrix model considered in this paper and
those considered in the above mentioned papers is the boundary
relation is not quadratic in $q$ (position) and $p$ (momentum), rather
it is trigonometric \cite{Dutta:2007ws, dutta-dutta} in general ,which
boils down to quadratic relation for small values of $\q$
(position). Another important difference is, in our case the free
fermions are living on group manifold which is compact, {\it i.e.}
position coordinate $\q$ is ranges between $[0, 2\pi]$ and momentum
$h$ is always positive.

The identification $\theta=\pi u(h)$ for $\theta>0$ is not 
proved in this article. 
Although this is the case we realize that the above has
correct bound on $u(h)$ and also is always positive. We intend 
to provide more rigourous arguments regarding this in our future 
work but here we present some heurestic arguments 
in appendix \ref{app:iden} which suggests the the identification is 
correct.

The second interesting result we present in this paper is a phase
space distribution of unfolded zeros of \rzf.  We also found a
similarity (or correspondence) between Young distribution function for
Riemann zeros and prime counting function. This observation gives a
strong hint that unfolded zeros of Riemann zeta function and prime
number behave like conjugate pairs, as also discussed in
\cite{Berry-Keating}. The eigenvalue distribution considered in this
paper has divergences at Riemann zeros, in stead one could consider a
matrix model whose eigenvalue density is exactly peaked at the
unfolded zeros {\it i.e.} $\r(\q)$ is equal to derivative of zero
counting function, {\it i.e.}
$$
\r(\q) \sim \frac d{d\theta} \lB\sum_{i=\text{zeros}}
\Theta(\q-\q_i)\rB \sim \sum_{i=\text{zeros}} \delta(\q-\q_i)
$$
and check the relation between Young tableaux distribution and prime
counting function. We hope to present the result in near future.

\bc
----------------------------
\ec

\noindent
{\bf Acknowledgement }

We would like to thank Rajesh Gopakumar for many useful discussion. SD
acknowledges the Simons Associateship, ICTP.  SD also thanks the
hospitality of ICTP, Trieste where part of this work has been done. We
thank Arghya Chattopadhyay for discussion. Finally, we are grateful to
people of India for their unconditional support towards researches in
basic sciences.

\input{appendix.tex}

\input{bib.tex}
\end{document}

%% file: gdefn.tex
\newcommand{\Tr}{\text{Tr}}

\newcommand{\av}[1]{\<{#1}\>}

\newcommand{\ben}{\begin{eqnarray}\displaystyle}
\newcommand{\een}{\end{eqnarray}}

\newcommand{\be}{\begin{equation}}
\newcommand{\ee}{\end{equation}}

\newcommand{\bc}{\begin{center}}
\newcommand{\ec}{\end{center}}

\newcommand{\eesp}{\end{split}}
\newcommand{\bsp}{\begin{split}}

\newcommand{\Rmnum}[1]{\expandafter\@slowromancap\romannumeral #1@}

\renewcommand{\a}{\alpha}	
\renewcommand{\b}{\beta}		
\newcommand{\dow}{\partial}
\newcommand{\e}{\epsilon}

\newcommand{\h}{\hbar}

\renewcommand{\l}{\lambda}	
\renewcommand{\o}{\omega}	
\newcommand{\q}{\theta}	
	
\renewcommand{\r}{\rho}		

\newcommand{\z}{\zeta}

\newcommand{\I}{\infty}

\renewcommand{\O}{\Omega}	

\newcommand{\cD}{\mathcal{D}}

\newcommand{\cN}{\mathcal{N}}

\newcommand{\cZ}{\mathcal{Z}}

\newcommand{\cota}[1]{\cot\lb #1 \rb}

\newcommand{\lna}[1]{\ln\lb #1 \rb}

\newcommand{\ra}{\rightarrow}

\newcommand{\lB}{\left [}
\newcommand{\rB}{\right ]}
\newcommand{\lb}{\left (}
\newcommand{\rb}{\right )}

\newcommand{\<}{\left\langle}
\renewcommand{\>}{\right\rangle}		

\newcommand{\rzf}{Riemann zeta function}
\newcommand{\zf}{ $zeta$ function}

\def\Xint#1{\mathchoice
{\XXint\displaystyle\textstyle{#1}}%
{\XXint\textstyle\scriptstyle{#1}}%
{\XXint\scriptstyle\scriptscriptstyle{#1}}%
{\XXint\scriptscriptstyle\scriptscriptstyle{#1}}%
\!\int}
\def\XXint#1#2#3{{\setbox0=\hbox{$#1{#2#3}{\int}$ }
\vcenter{\hbox{$#2#3$ }}\kern-.6\wd0}}

%% file: plaq.tex
\section{Single plaquette model and properties of 
resolvent}
\label{sec:plaq}

Unitary matrix model of the following form has great importance in
different branches of physics. Different $2d$ gauge theories,
supersymmetric gauge theories in various dimensions can be cast in
terms of unitary matrix models. These models also contain a rich phase
structure.  In this paper we consider a particular class of unitary
matrix model, called single plaquette model. However, our results are
also valid for a class of model considered in \cite{shiraz,
  AlvarezGaume:2005fv, AlvarezGaume:2006jg, Aharony:2005ew,
  Aharony:2005bq, Sundborg:1999ue, Yamada:2006rx, Grassi:2014vwa,
  Marino:2012zq, Marino:2011nm, Aganagic:2002wv} up to some
re-definitions of parameters.

We consider the following partition function over $N\times N$ unitary
matrices $U$
\begin{eqnarray}\label{eq:PF0}
Z=\int {\cal{D}}U\, \exp\left(NTr[f(U)]+NTr[f(U^{\dagger})]\right).
\end{eqnarray}
Moreover let us assume that the function $f(z)$ has a convergent
Taylor series expansion
\begin{eqnarray}
f(U)&=&  \sum_{n=0}^{\infty}\frac{f^{n}(0)}{n{!}}U^{n}\nonumber\\
&=&\sum_{n=0}^{\infty}\frac{\b_n}{n}U^{n}, 
\qquad \b_n = \frac{f^{n}(0)}{(n-1){!}}.
\end{eqnarray}
Hence, the partition function can be written as,
\begin{eqnarray}
Z=\int {\cal{D}}U\, \exp(N\sum_{n=0}^{\infty}
\frac{\b_{n}}{n}(Tr[U^{n}]+Tr[U^{\dagger n}])).
\end{eqnarray}
This is called a single plaquette model. $\beta_n$'s are parameters of
this theory. They can depend on temperature of the system (for example
\cite{shiraz, gross-witten, wadia}). Hence, as temperature changes,
$\b_n$'s change and the system undergoes phase transition form one
phase to another phase.

\subsection{Eigenvalue analysis}

An eigenvalue analysis of this model has been discussed in
\cite{Jurkiewicz:1982iz}\footnote{See \cite{Marino:2004eq} for a
  review lecture on matrix model.}.  Since integrand and Haar measure
$\cD U$ are invariant under unitary transformation, one can go to a
diagonal basis where $U$ has eigenvalues $\q_i,\ i = 1, \cdots, N$ and
write the partition function as,
\be\label{eq:pfeige}
\cZ = \int \prod_{i=1}^N d\q_i e^{S[\q_i]}
\ee
where,
\be
S[\q_i] = N \sum_{n=1}^{\infty} \sum_{i=1}^N {2\b_n\over n}
\cos n\q_i +\frac12 \sum_{i\neq j} 
\ln \left|4 \sin^2 \lb{\q_i-\q_j\over 2}\rb\right|.
\ee
In the large $N$ limit, one can define continuous variables
\be
x={i\over N} \in[0,1],\qquad \q(x) =\q_i 
\ee
and the partition function is given by,
\ben
\label{eq:pfeige2}
\begin{split}
\cZ &= \int [\cD \q] e^{-N^2 S[\q]}, \ \ \text{where}\\
S[\q] &= \sum_{n=1}^{\infty} \int_0^1 dx {2\b_n\over n}
\cos n\q(x) + \frac12 \int_0^1dx\Xint-_0^1 dy 
\ln \left|4 \sin^2 \lb{\q(x)-\q(y)\over 2}\rb\right|.
\end{split}
\een
In the large $N$ limit, the dominant contribution comes from extrema
of $S[\q]$ and that is determined by saddle point equation,
\be\label{eq:eveqnplaq} \Xint-_{-\pi}^{\pi} d\theta'\,\rho(\theta')
\cota{\frac{\q-\theta'}{2}} = V'(\q), \ee
where,
\be\label{eq:V'def}
\r(\q) = {\dow x\over \dow \q} \quad 
\text{and} \quad V'(\q)=\sum_{n=1}^{\infty}2\b_n\sin n\q.
\ee
Thus for a given set of $\b_n$'s one has to solve saddle point
equation (\ref{eq:eveqnplaq}) to find $\r(\q)$.

\subsection{The resolvent and its properties}
\label{sec:Rzprop}

It is, in general, difficult to solve an integral equation to find
eigenvalue distribution. In this section we discuss an equivalent
approach to study phases structure of unitary matrix models from
analytic properties of $resolvent$.

The resolvent is defined by
\be \label{eq:Rz}
R(z) =N^{-1}\langle Tr[(1-zU)^{-1}]\rangle.
\ee
Expanding the right hand side one can write,
\be\label{rzexpan}
R(z) = 1 + \frac1N \lb z \<\Tr U\> + z^2  \<\Tr U^2\> + z^3 \<\Tr
U^3\> +\cdots \rb.
\ee
Since $N^{-1}\av{\Tr U^k}$ lies between $-1$ and $+1$ the right hand
side of eqn. (\ref{rzexpan}) is convergent for $|z|<1$. Therefore,
$R(z)$ is an analytic and holomorphic function of $z$ in the interior
of unit disk.

Analyticity of resolvent outside the unit disk can also be shown
easily. Expanding the function around $z=\infty$ we find,
\be \label{rzexpan2}
R(z) = - \frac1N \lb \frac1z \<\Tr U\> + {1\over z^2}  \<\Tr U^2\> 
+ {1\over z^3} \<\Tr
U^3\> +\cdots \rb.
\ee
The series converges for $|z|>1$. From equation (\ref{rzexpan}) and
(\ref{rzexpan2}) we find that the resolvent satisfies,
\be\label{eq:Rzprop1}
R(z)+R(1/z)=1.
\ee

The advantage of this approach is that in large $N$ limit resolvent
satisfies an algebraic equation (quadratic) rather than an integral
equation and hence easy to solve. The equation can be obtained from
the fact that Haar measure is invariant under a generalized unitary
transformation $U\ra U e^{t X}$, where $X$ is a skew symmetric matrix
and $t$ is a real parameter. Partition function (\ref{eq:PF0}) being
invariant under such transformation yields the following identity,
which is known as Dyson-Schwinger equation
\begin{eqnarray}
\begin{split}
\frac{d}{dt}\int {\cal{D}}U\, N^{-1}Tr[(1-zUe^{tX})]\exp(N
(Tr[(f(Ue^{tX}))]+Tr[f(U^{\dagger}e^{-tX})])&=0. \\
\frac{d}{dt}\int {\cal{D}}U\, N^{-1}Tr[(1-zUe^{tX})]\exp(N\sum_{n=0}^{\infty}
\frac{\b^{n}}{n}(Tr[(Ue^{tX})^{n}]+Tr[(U^{\dagger}e^{-tX})^{n}]))&=0.
\end{split}
\end{eqnarray}
A solution to this equation is given by,
\begin{eqnarray}
\begin{split}
R(z)&=\frac{1}{2}[1+(f'(z)z-f'(1/z)/z)+\sqrt{F(z)}]\quad\quad |z|<1\\
R(z)&=\frac{1}{2}[1+(f'(z)z-f'(1/z)/z)-\sqrt{F(z)}]\quad\quad |z|>1
\end{split}
\end{eqnarray}
where, $F(z)$ is invariant under $z\to 1/z$.

Expanding $f(z)$ in Taylor series we find the solution is given by,
\begin{eqnarray}
\begin{split}
R(z) &= \frac{1}{2}\lB1+\sum_{n=1}^{m}\beta_{n}(z^{n}-
\frac{1}{z^{n}})+\sqrt{F(z)}\rB\quad\quad |z|<1\\
R(z) &= \frac{1}{2}\lB 1+\sum_{n=1}^{m}\beta_{n}(z^{n}-
\frac{1}{z^{n}})-\sqrt{F(z)}\rB\quad\quad |z|>1
\end{split}
\end{eqnarray}
and $F(z)$ has the following form
\begin{eqnarray}
&&F(z)=\bigg[1+\sum_{i=1}^{m}\beta_{i}[\frac{1}{z^{i}}+z^{i}]\bigg]^{2}
-4\bigg[\sum_{i=1}^{m}\frac{\beta_{i}}{z^{i}}\bigg]\bigg[\sum_{i=1}^{m}
\beta_{i}z^{i}\bigg]+4\beta_{1}R'(0)\notag\\
&&+4\beta_{2}\bigg[\frac{R'(0)}{z}+\frac{R''(0)}{2}+R'(0)z\bigg]+4\beta_{3}
\bigg[\frac{R'(0)}{z^{2}}+\frac{R''(0)}{2z}+\frac{R'''(0)}{3{!}}
+\frac{R''(0)z}{2}+R'(0)z^{2}\bigg]+\cdots\notag
\end{eqnarray}
Analyticity properties of $R(z)$ inside a unit circle depends on the 
form of $F(z)$ which determines phase structure of this model. A 
detailed discussion can be found in \cite{dutta-dutta}.

\subsection{Eigenvalue distribution function from 
resolvent}

The eigenvalue density can be obtained from real part of the resolvent
\cite{Friedan:1980tu},
\be\label{eq:rhodef}
\rho(\theta) = 2\Re [R(e^{i\theta})]-1
\ee
which can also be written as,
\be
\rho(\theta) = \lim_{\epsilon\ra 0} \frac1{2\pi}\lB R((1+\e)e^{i\theta})
-R((1-\e)e^{i\theta}) \rB.
\ee
On the other hand, imaginary part of $R(z)$ gives the derivative of 
potential $V'(\theta)$ which appears on the r.h.s. of 
saddle point equation (\ref{eq:eveqnplaq})
\be \label{eq:V'def2}
V'(\theta) = 2\Im[R(e^{i\theta})].
\ee

%% file: rzf.tex
\section{Riemann zeta function and its properties}
\label{sec:rzf}

In $18^{th}$ century famous mathematician Leonhard Euler defined a
function denoted by $\zeta$,
\be\label{rzf1}
\zeta(s) = \sum_{n=1}^{\infty} {1\over n^s}
\ee
for positive integer values of $s>1$. Later, famous Russian
mathematician Chebyshev extended the definition for any real values of
$s > 1$.

In 1859 Bernhard Riemann extended zeta function defined by Euler to
complex variable and showed that this function can be analytically
continued to whole complex $s$ plane except the point $s=1$. Thus the
function defined in equation (\ref{rzf1}) as a function of complex
variable $s$ is know as \rzf \ (or Euler-\rzf) and is a meromorphic
function of complex variable $s$. See \cite{riemann, hmedwards} for
detail discussion on \rzf.

\subsection{Poles and zeros of $zeta$ function and
Riemann Hypothesis}

\rzf \ has a simple pole at $s=1$. It has infinite number of zeros at
$s=-2n$ for positive integer $n$. These zeros are called
$trivial \ zeros$ since their distribution is known. The function also
has infinitely many zeros, known as $non-trivial\ zeros$ (or unfolded
zeros), in the critical strip $0<\Re [s] <1$. Distribution of these
non-trivial zeros are not known completely.

Riemann Hypothesis states that all on-trivial zeros lie on the
critical line $\Re [s] =1/2$. This means,
$$
\zeta\lb\frac12+i t_i\rb =0
$$
has non-trivial solution for $t_i \in \mathbb R$. Riemann Hypothesis
is not yet proved and is known to be true for at least 40 $\%$ of
non-trivial zeros.

 \subsection{Few important properties of $zeta$ function}

\subsubsection{Functional equation and symmetric $zeta$ function}

Riemann zeta function satisfies the following functional equation
\be \label{eq:functionaleq}
\zeta(s) = 2^s \pi^{s-1}\sin\lb {\pi s\over 2} \rb
\Gamma(1-s)\zeta(1-s).
\ee
Here $\Gamma(s)$ is gamma function. This equation relates value of
zeta function at two different points in complex plane. Due to
presence of $\sin$ function on the right hand side, it is easy to
check that zeta function vanishes at $s=-2|n|$ for any integer. For
$s= +2|n|$, the Gamma function has simple poles at non-positive
integers and hence the product $\sin(\pi s/2)\Gamma(1-s)$ on the right
is regular and non-zero.

Riemann also defined a symmetric version of $zeta$ function, denoted by, 
$\xi(s)$,
\be\label{xifunc}
\xi(s)= \frac12 \pi^{-s/2} s(s-1) \Gamma\lb \frac s2\rb \zeta(s).
\ee
Zeta function has zeros at $s=-2|n|$ whereas $\Gamma$ function has 
poles at those points, hence $\Gamma\lb \frac s2\rb \zeta(s)$ is regular at 
$s=-2|n|$.
Since $\zeta(s)$ has a pole order one at $s=1$ with residue 1,
$\xi(s)$ has a normalization $\ln\xi(1)=-\ln 2$. Therefore,
the symmetric version of zeta function $\xi(s)$ is analytic 
everywhere in complex $s$ plane. From functional 
equation (\ref{eq:functionaleq}) one can also show that, 
symmetric zeta function $\xi(s)$ satisfies
\be\label{eq:synzetaeqn}
\xi(s) =\xi(1-s).
\ee
In our paper we shall mostly consider this symmetric $zeta$ function.

\subsubsection{Product formula}
 
 There is a strong connection between zeta function and 
 prime numbers. Euler proved that there exists a product
 representation of $zeta$ function given by,
 \be\label{eq:zetaprod}
 \zeta(s) = \prod_{p= prime} {1\over 1-p^{-s}}.
 \ee
 The symmetric $zeta$ function also satisfies following 
 product representation
 \begin{eqnarray} \label{eq:xiprod}
\xi(s) =\frac12 \prod_{i=1}^{\infty}\bigg(1-\frac{s}{\rho_{i}}\bigg)
\end{eqnarray}
where $\rho_{i}$ are  non trivial zeroes of $\zeta$
function.

 \subsubsection{Li criteria}

 In 1997, Xian-Jin Li proved that the Riemann hypothesis for zeta
 function is equivalent to non-negativity of a sequence of real
 numbers \cite{Li}\footnote{See also \cite{Bombieri}}.  Li introduced
 following sequence of numbers
\begin{eqnarray}\label{eq:Linumbers}
&&(n-1){!}\lambda_{n}=\frac{d^{n}}{ds^{n}}\lB s^{n-1}\ln\xi(s)\rB_{s=1}
\end{eqnarray}
and showed that a necessary and sufficient condition for  nontrivial zeros
of Riemann \zf \ to lie on critical line is that $\lambda_n$ 
is non-negative for every positive integer $n$. 
 
Using product representation (\ref{eq:xiprod}) of symmetric zeta function
one can show that the Li numbers can be written as,
\begin{eqnarray}\label{eq:Linumber}
\lambda_{n}=\sum_{i}\bigg[1-\bigg(1-\frac{1}{\rho_{i}}\bigg)^{n}\bigg].
\end{eqnarray}

%% file: model.tex
\section{The matrix model} 
\label{sec:model}

Having discussed relevant properties of plaquette model, resolvent and
zeta function, we are now in a position to construct a unitary matrix
model whose eigenvalue distribution captures information about zeros
of \rzf.

\subsection{From $s-$plane to $z-$plane}

\begin{figure}[h]
\begin{subfigure}{.5\textwidth}
\centering
\includegraphics[width=7cm,height=7cm]{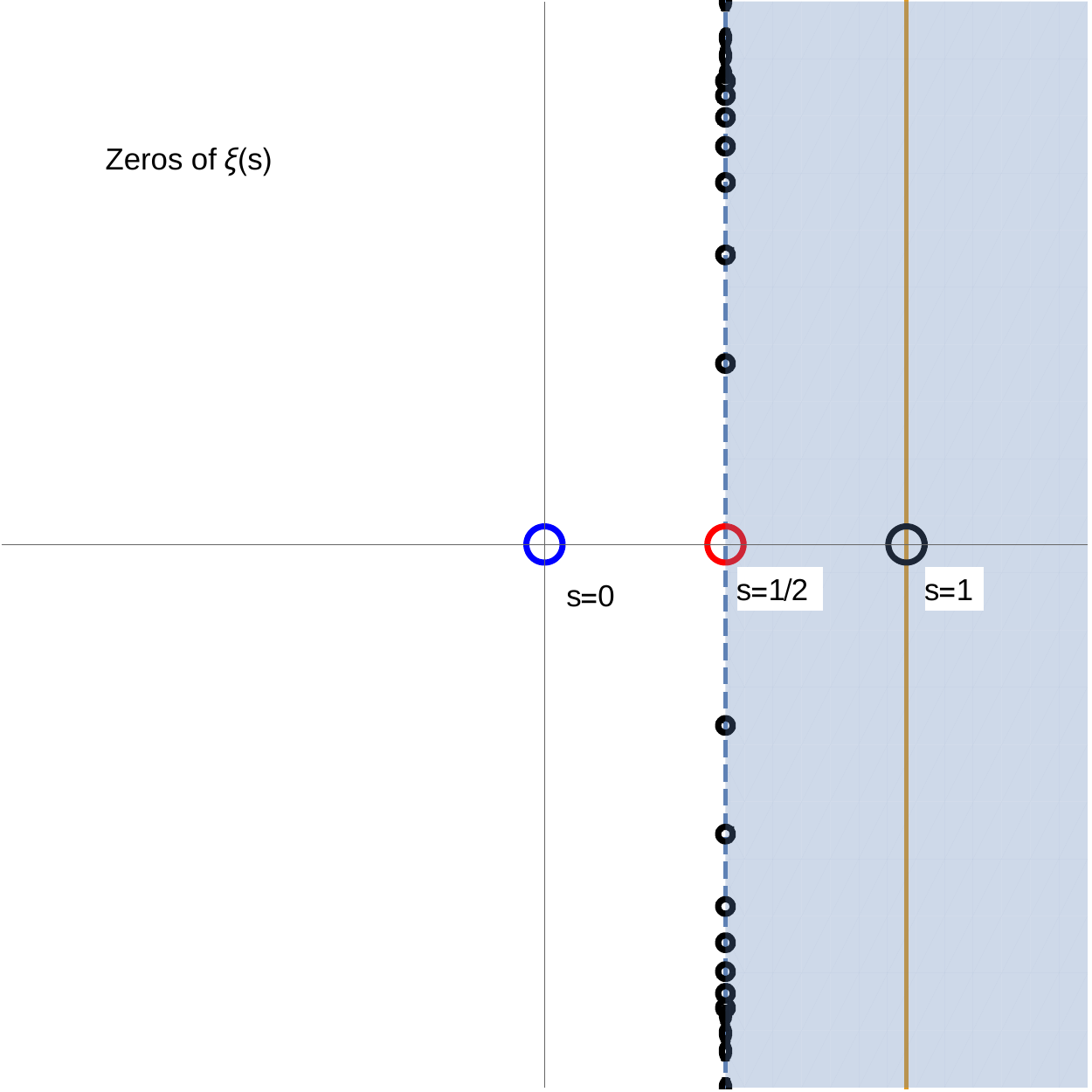}
\caption{}
\end{subfigure}%
\begin{subfigure}{.5\textwidth}
\centering
\includegraphics[width=7cm,height=7cm]{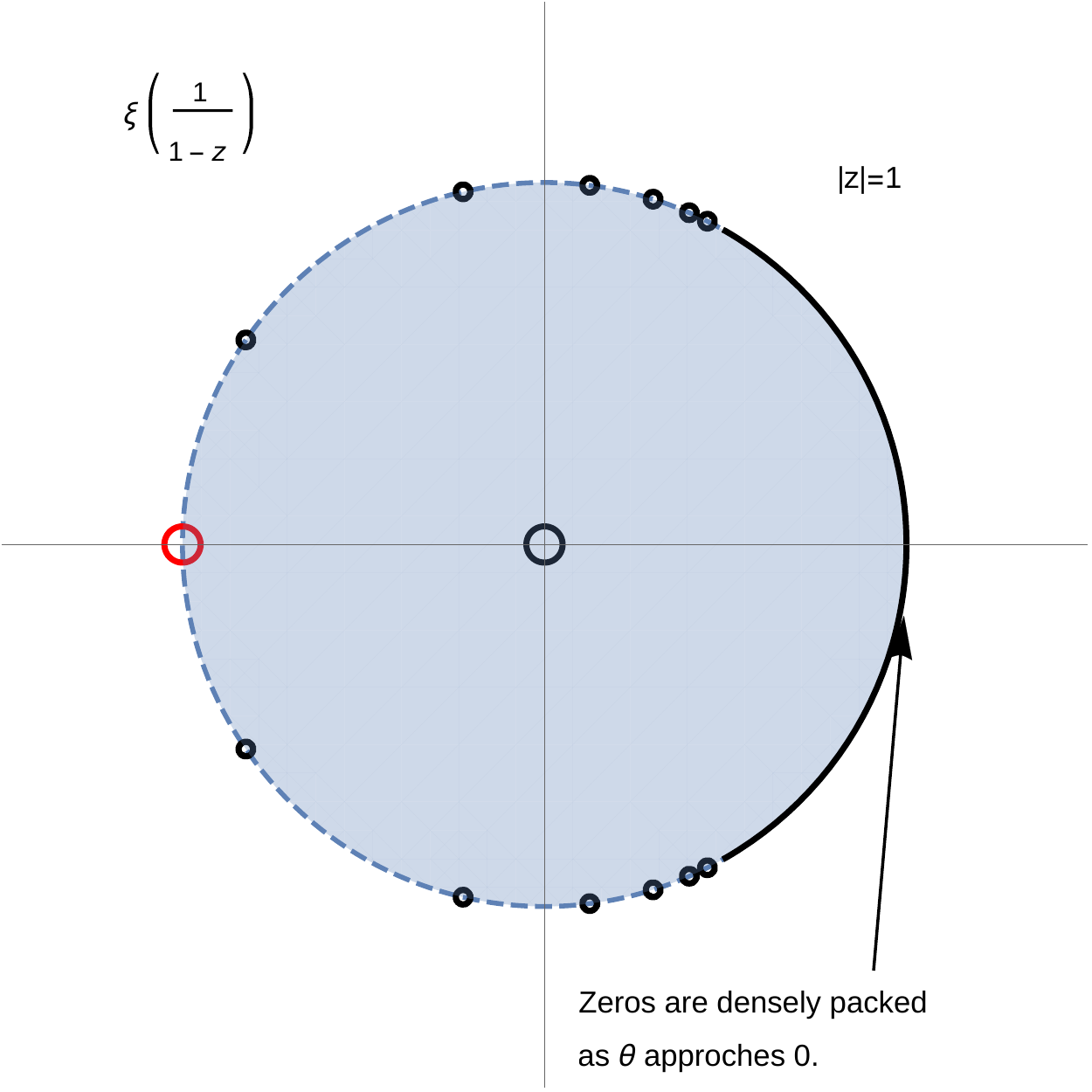}
\caption{}
\end{subfigure}
\caption{$s$ plane vs. $z$ plane and unfolded zeros of
\rzf}
\label{fig:s-zmapp}
\end{figure}

Since eigenvalues of unitary matrices are distributed on a unit circle
in complex $z$ plane, we need to first map the conjectured zeros (on
critical line in $s$ plane) of zeta function on a unit circle about
the origin in $z$ plane.  We use the following conformal map
\be\label{eq:confmap}
s= {1\over 1-z}.
\ee
All unfolded zeros on critical line (assuming Riemann hypothesis) in
$s-$plane corresponds to $s=1/2+i t_i$ where $t_i \in \mathbb R$.  In
$z-$plane these zeros correspond to $z= e^{i\theta_i}$. Hence, using
equation (\ref{eq:confmap}) we find,
\be
t_i = \frac12 \cot{\theta_i \over 2}.
\ee
The mapping is explained in figure \ref{fig:s-zmapp}.  The critical
line $\Re [s] =1/2$ in $s$ plane is mapped to a unit circle about origin
in $z$ plane (dashed line in figure \ref{fig:s-zmapp}).  The region
$\Re[ s] >1/2$ in $s$ plane is mapped inside the unit circle in $z$
plane (shaded region) and region $\Re [s] <1/2$ is mapped outside the
unit circle. $s=1$ point is mapped to $z=0$ point, $s=1/2$ is mapped
to $z=-1$.

Since symmetric zeta function (\ref{xifunc}) is analytic in $s$ plane,
it is therefore analytic inside and outside the unit circle in $z$
plane with unfolded zeros located on $|z|=1$. In $s-$plane zeros of
\rzf\ are symmetrically distributed about $\Im [s] =0$, this implies
in $z-$plane zeros are symmetrically distributed on unit circle about
$\Im[z]=0$. It is observed that distribution of zeros of \zf \ becomes
denser as $t_i$ becomes larger. Therefore, in $z-$ plane density of
distribution increases as one goes to $\theta \ra 0$.

Functional equation (\ref{eq:synzetaeqn}) for $\xi$ in $z-$plane is
given by,
\be
\xi\lb {1\over 1-z}\rb = \xi\lb {-z\over 1-z}\rb.
\ee
%

\subsection{The resolvent}
\label{sec:resolvent}

As discussed in section \ref{sec:plaq}, resolvent contains all the
information about matrix model and eigenvalue distribution of its
different phases. Eigenvalue distribution is given by real part of
resolvent defined by equation (\ref{eq:Rz}), whereas imaginary part of
resolvent contains information about potential.

Given a unitary matrix model, one needs to solve the Dyson-Schwinger
equation to find resolvent with correct analytic structure. Eigenvalue
distribution can be obtained from real part of the resolvent. Here we
follow an inverse route. From the analytic properties, we construct a
resolvent which captures information of unfolded Riemann zeros. Then,
from resolvent we write down a consistent unitary matrix model such
that the resolvent satisfies corresponding Dyson-Schwinger equation.

Following the properties of resolvent given in section
\ref{sec:Rzprop}, we take following form of the resolvent
\begin{equation}
\begin{split}\label{eq:Rzinout}
R_<(z)& =\frac{1}{2}-\alpha\ln\xi\lb\frac{1}{1-z}\rb\quad\quad|z|<1 \\
R_>(z)& =\frac{1}{2}+\alpha\ln\xi\lb \frac{1}{1-z}\rb\quad\quad|z|>1.
\end{split}
\end{equation}
We fix value of $\a$ from boundary condition $R(0)=1$,
which implies
\be
\a = -{1\over 2 \ln \xi(1)} = {1\over 2 \ln 2}.
\ee
The above form of $R(z)$ satisfies all the properties. First of all
$R_<(z)$ is analytic inside unit circle (assuming Riemann
hypothesis). Hence there exists a Taylor expansion around $z=0$ inside
unit circle. Symmetric properties of $\xi(z)$ ensures condition
(\ref{eq:Rzprop1}),
\be
R_<(z)+R_>(1/z) = 1 + \a \ln \lB {\xi\lb{1\over 1-z} \rb 
\over \xi\lb z\over z-1\rb}\rB =1.
\ee
Since $\xi({1\over 1-z})$ has nontrivial zeros on unit circle in
$z-$plane, $\ln\xi({1\over 1-z})$ has logarithm branch points on unit
circles.

Since $\xi(z)$ is analytic in $z$ plane, there are different
possibilities to choose a resolvent which satisfies required
properties. In the next section we see that this particular choice of
resolvent gives rise to no-gap solution {\it i.e.} eigenvalue
distribution has no gap. Also, corresponding unitary model turns out
to be simple. This particular choice also ensures that corresponding
Young distribution has a direct connection with prime counting
function.

\subsection{The eigenvalue density}

The eigenvalue distribution according to equation (\ref{eq:rhodef}) is
given by,
\ben
\r(\theta) = -2\alpha\Re\lB\ln\xi\lb \frac{1}{1-z}\rb
\bigg|_{z=e^{i\theta}}\rB.
\een
Our choice of resolvent indicates that unfolded zeros \rzf \ will play
special role in $\rho(\theta)$.

Using product representation (equation \ref{eq:xiprod}) of symmetric
zeta function and denoting the positions of unfolded zeros by $\q_i$
we write a complete expression for eigenvalue distribution as (see
appendix \ref{app:evcalcu1} for detailed calculation),
\ben\label{eq:evdensity}
\r(\theta) = 1 + 2\alpha \sum_i \ln \lb \sin{\theta\over 2}\rb
-\a \sum_i \ln \lB \sin^2\lb {\theta-\theta_i \over 2} \rb\rB.
\een
%
\begin{figure}[h]
\centering
\includegraphics[width=10cm,height=6cm]{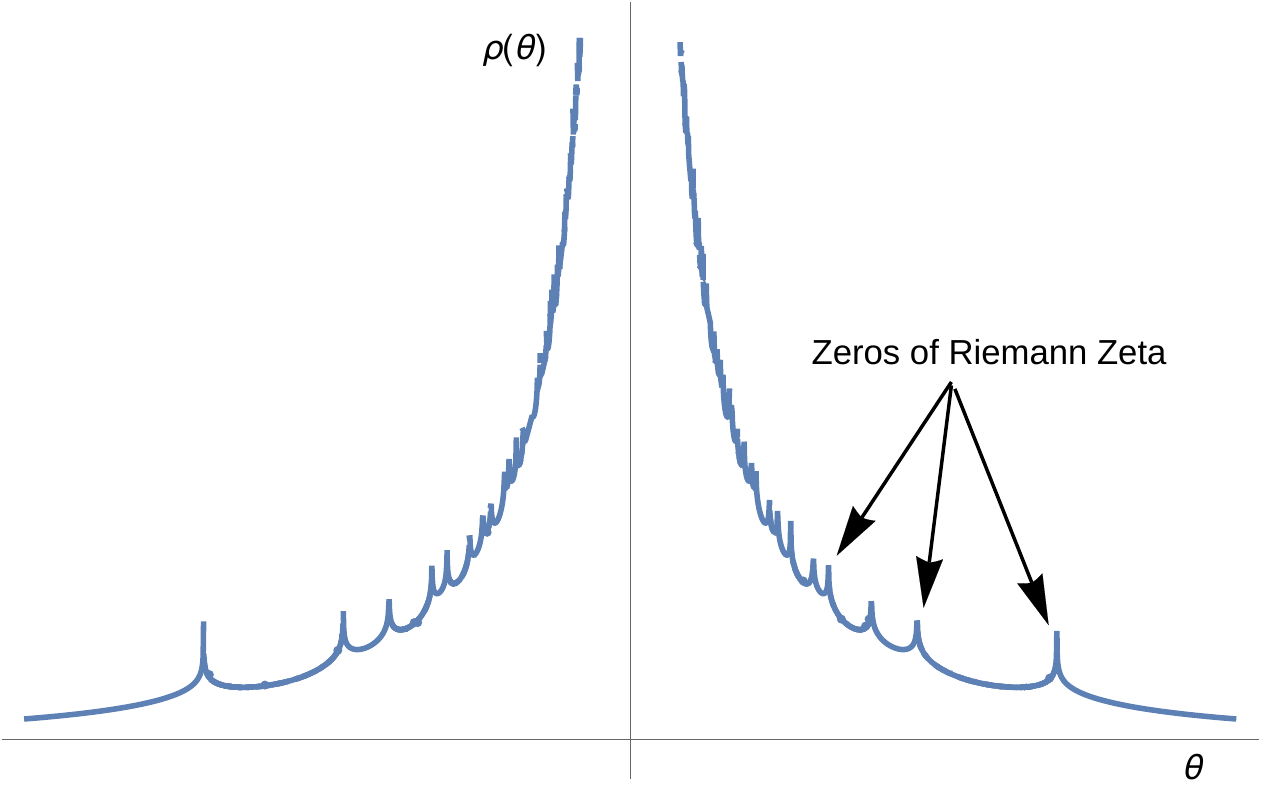}
\caption{Eigenvalue distribution. We see that 
the distribution has logarthimic divergences
at Rieman zeros.}
\label{fig:rhoplot}
\end{figure}

Since $|\q_i|<\pi$ for all $i$, we see that from the last term
$\r(\q)$ diverges logarthimically at $\q_i$'s. Therefore, this 
term in $\r(\q)$ actually carries information about distribution of
Riemann zeros. The second term also diverges at $\q=0$, but this is 
usual divergence as $t\ra \pm \infty$ in complex $s$ plane. Although,
we see logarthimic divergences in $\r(\q)$, however, it is easy to check 
that the eigenvalue density satisfies normalization condition
\ben
\frac{1}{2\pi}\int_{-\pi}^{\pi} \r(\theta)d\theta &=& 1. 
\een
See appendix \ref{app:rhonormalization} for details. 

In figure \ref{fig:rhoplot} we plot 
$\r(\q)$ as a function of $\q$. Sharp peaks correspond to logarthimic 
divergences at Riemann zeros. Since Riemann zeros are densely packed
around $\q=0$, therefore most of the peaks appear near $\q=0$.

Eigenvalue distribution given by equation (\ref{eq:evdensity}) corresponds
to a no-gap solution because, the distribution does not have any 
gap between $-\pi<\q\leq\pi$.

\subsection{The plaquette model}

In this section we construct a single plaquette model whose
eigenvalue distribution exactly matches with equation (\ref{eq:evdensity}).

A single placquette model action is given by,
\begin{eqnarray} \label{eq:placquette}
S(U)=N \sum_{n=1}^{\infty}{\beta_{n}\over n}
(Tr[U^{n}]+Tr[U^{\dagger n}])
\end{eqnarray}
where $\b_n$'s are constant. Eigenvalue distribution in different
phases of this model depends on values of $\b_n$'s.

Since resolvent $R(z)$ is analytic inside unit circle one can write
down a Taylor expansion of $R(z)$ about $z=0$ for $|z|<1$
\be\label{eq:Rzin}
R(z) = 1+ \sum_{n=1}^{\infty} \a_n z^n, \quad 
\text{for} \quad |z|<1
\ee
with $\a_n = R^{(n)}(0)/n!$ and $R(0)=1$. Plugging this expression
in Dyson-Schwinger equation and demanding that the equation is
satisfied inside unit circle one can find $\a_n$ in terms of
$\b_n$'s for different phases. However, for no-gap phase there exists
a trivial solution to Dyson-Schwinger equation : $\a_n=\b_n$.

Since, we are trying to construct a plaquette model with a
a no-gap phase, we have
\ben
\a_n=\b_n = \frac1{n!} R^{(n)}(0)={1\over 2\pi i} \oint {R(z)\over 
z^{n+1}} dz,
\een
where the contour is takes around $z=0$. 

Recall definition of Li numbers given by equation (\ref{eq:Linumbers}).
These numbers can also be written in integral form using
Cauchy theorem,
\begin{eqnarray}
(n-1){!}\lambda_{n}=\frac{n{!}}{2\pi i}\oint_{c}ds
\frac{s^{n-1}}{(s-1)^{n+1}}\ln\xi(s),
\end{eqnarray}
where the contour $c$ is taken counter clockwise around $s=1$. In 
$z-$plane the Li numbers are given by, 
\begin{eqnarray}
\frac{\lambda_{n}}{n}=\frac{1}{2\pi i}\oint_{c}\frac{dz}
{z^{n+1}}\ln\xi\lb\frac{1}{1-z}\rb.
\end{eqnarray}
Using the explicit form of $R(z)$ inside unit circle (equation 
\ref{eq:Rzinout}), one can write
\begin{eqnarray}
\a \frac{\lambda_{n}}{n}= -{1\over 2\pi i} \oint {R(z)\over 
z^{n+1}} dz
\end{eqnarray}

Hence, if we take 
\be
\beta_n = - \a {\lambda_n \over n}, \ \ \text{for all}\ n
\ee
then the above series (\ref{eq:Rzin}) can be re-summed and written as
\be
R_<(z) =\frac{1}{2}-\alpha\ln\xi\lb\frac{1}{1-z}\rb\quad\quad|z|<1 .
\ee
Thus we get a placquette model whose eigenvalue 
distribution for a particular phase is given by equation (\ref{eq:evdensity}) 
and captures information about distribution of non-trivial
zeros of \rzf. The model is given by,
\begin{eqnarray}\label{eq:plaqmodlzeta}
S(U)=-\alpha\sum_{n=1}^{\infty}\frac{\lambda_{n}}{n^2}
(Tr[U^{n}]+Tr[U^{\dagger n}]).
\end{eqnarray}
Since real part of resolvent contributes to eigenvalue
distribution, one can compute from equation (\ref{eq:Rzin}) 
that eigenvalue distribution
is given by,
\begin{eqnarray}
\r(\theta)=\frac{1}{2\pi}\bigg(1+2\alpha
\sum_{n=1}^{\infty}\frac{\lambda_{n}}{n}\cos n\theta\bigg).
\end{eqnarray}
Upon substituting values of $\l_n$ given in equation (\ref{eq:Linumber})
\begin{eqnarray}
\lambda_{n}=\sum_{i}\bigg[1-\bigg(1-\frac{1}{\rho_{i}}\bigg)^{n}\bigg]
\end{eqnarray}
one can show that eigenvalue density can be written as
(\ref{eq:evdensity}).

\subsection{The potential}

We have constructed the matrix model from real part of the resolvent.
As a consitency check we shall construct the potential from imaginary
part of resolvent and compare with equation (\ref{eq:plaqmodlzeta}).

Derivative of the potential is given by imaginary part of the
resolvent. Since $\xi$ function has a product representation give by
equation (\ref{eq:xiprod}), therefore $\Im[\ln \xi(s)]$ picks up a
$i\pi$ factor everytime we cross a non-trivial zero $\rho_i$. We can
write,
\ben
\begin{split}
\ln \xi(s) &= \sum_{i} \ln \lb 1- {s\over \r_i}\rb\\
&= \sum_{i} \ln \lb \r_i- s\rb -\sum_i \ln \r_i,
\end{split}
\een
where $i$ runs over all non-trivial zeros in upper and lower half
plane. Denoting $\r_i = 1/2+i\ r_i$ and $s=1/2+i \ t$ for
$r_i,\ t \in \mathbb{R}$ we can write
\be \label{eq:logxis}
\ln \xi(s) =  \Sigma^U_i\ln (r_i-t) + \Sigma_i^{L} 
\ln(t-r_i) -\Sigma_i \ln \r_i
\ee
where $\Sigma^{U/L}$ implies sum over non-trivial zeros in
upper/lower half plane.

Since non-trivial zeros are distributed symmetrically
about the real axis, one can show that $\sum_i \ln \r_i$
is a real number. Only the first two terms contribute 
to imaginary part of $\ln \xi(s)$. Every time one 
crocess a non-trivial zero in upper half plane $r_i-t$
becomes negative and the first logarithm  
 gives an extra $i \pi$ term, i.e.
\be
\ln(r_i-t) \ra \ln(t-r_i) +i \pi\quad \text{for} \quad 
t>r_i.
\ee
Note that, in lower half plane all $r_i$'s are negative 
and hence second term in equation (\ref{eq:logxis}) is always
positive for $t>0$.
Therefore imaginary part of $\ln \xi(s)$ is a step
function
\be
\Im\lB \ln(s) \rB = \Sigma_i^U \Theta(t-r_i) \quad \text{for}
\quad t>0
\ee
The story is same for $t<0$ also. This time  second logarthm gives
an imaginary part as one crocess a non-trivial zero in lower
half plane. However, for lower zeros we shall get an extra phase factor,
i.e., when we cross $i^{th}$ zero
\be
\ln (t-r_i)=\ln(|t|-|r_i|)-i\pi.
\ee
Therefore, the final result is
\ben
\begin{split}
\Im\lB \ln(s) \rB &= \ \ \Sigma_i^U \Theta(t-r_i) \quad \text{for}
\quad t>0 \\
&=-\Sigma_i^L \Theta(r_i-t) \quad \text{for}
\quad t<0
\end{split}
\een

\begin{figure}[h]
\centering
\includegraphics[width=10cm,height=6cm]{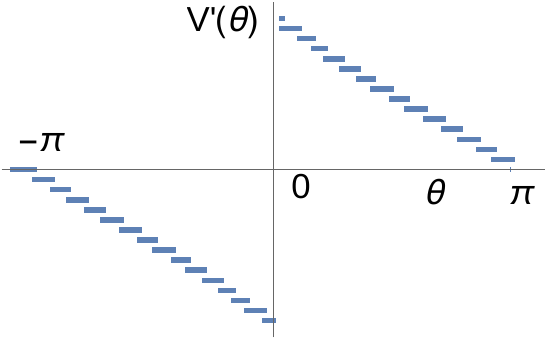}
\caption{Derivative of potential. We see that 
the derivative of potential has discrete jumps
at Rieman zeros.}
\label{fig:potplot}
\end{figure}

Hence, in $z-$plane imaginary part of 
$R(z=e^{i\theta})$ can be written as,
\ben\label{eq:imrz}
\begin{split}
\Im \lB\ln \xi\lb {1\over 1-e^{i \theta}}\rb\rB
&= - \sum_{\text{zeros}} \Theta(\theta-\theta_i),\quad -\pi\leq \theta\leq 0 \\
&= \ \ \ \sum_{\text{zeros}} \Theta(\theta_i-\theta) ,\quad 0 < \theta \leq \pi.
\end{split}
\een
Thus derivative of the potential is biven by (see figure \ref{fig:potplot})
\ben
\begin{split}
V'(\theta)
&= -\a \sum_{\text{zeros}} \Theta(\theta-\theta_i),\quad -\pi\leq \theta\leq 0 \\
&= \a \sum_{\text{zeros}} \Theta(\theta_i-\theta) ,\quad 0 < \theta \leq \pi.
\end{split}
\een
%

From equation (\ref{eq:Rzin}), imaginary part of resolvent can 
also be written as
\be
\Im \lB\ln \xi\lb {1\over 1-e^{i \theta}}\rb\rB= 
-{\a}\sum_n{\l_n \over n}\sin n\q.
\ee
Again, plugging values of $\l_n$ (equation \ref{eq:evdensity}) one 
can show that imaginary part of resolvent matches with equation
(\ref{eq:imrz}).

\subsection{An alternate proof of Li theorem}

One can prove Li theorem, i.e., positivity of Li numbers $\l_n$ from
analytic properties of resolvent which satisfies Dyson-Schwinger equation
of an underlying unitary matrix model.

The resolvent on unit circle is given by,
\ben
\begin{split}
R(z &=e^{i\theta}) = 1 + \sum_n \beta_n z^n\\
&= 1-\a  \sum_n {\l_n\over n} \cos n\theta 
- i \a  \sum_n {\l_n\over n} \sin n \theta.
\end{split}
\een
Therefore Li numbers $\l_n$'s can be obtained either from 
real or imaginary part of resolvent. Writing, Li numbers in 
terms of imaginary part we find,
\be
-\a {\l_n\over n} = \int_{-\pi}^{\pi} \Im\lB R(z=e^{i\theta})\rB
\sin n\theta d\theta.
\ee
Using equation (\ref{eq:imrz}) and doing integration by parts 
one can show that,
\be
\l_n = \sum_i\lb 1- \cos n \theta_i\rb \geq 0.
\ee
This proves Li's theorem. This proof is based on the fact that there
exists a unitary model, whose eigenvalue density is given by equation
(\ref{eq:evdensity}) and corresponding resolvent satisfies
Dyson-Schwinger equation which implies $\b_n = -\a \l_n/n$.

%% file: young.tex
\section{Generalised phasespace distribution of unitary matrix model}
\label{sec:Young}

Possibility of free fermi phase space description of large N saddle
points of a unitary matrix model was first observed in
\cite{douglas}. This was possible because partition can also be
written as sum over representation of underlying $U(N)$ group. Since
different representations can be described in terms of Young diagram,
it was shown in \cite{khazakov} that in large $N$ limit partition
function is dominated by class of Young diagrams. It turns out that,
there exists a relation between the most dominant representation of
$U(N)$ group and eigenvalue distribution which enables one to define a
free fermi phase space distribution considering eigenvalues $\q_i$ of
$U(N)$ as coordinates of fermions and number of boxes $h_i$ in
$i^{th}$ row of a Young diagram as conjugate momentum. This alternate
possibility of writing partition function enables one to give a
topological or geometrical description of large $N$ saddle points in
terms of phase space distribution.

The difficulty in understanding universal nature of free fermi
description lies in finding an exact expression of character of
permutation group in presence of arbitrary number of cycles. In
\cite{Dutta:2007ws} simpler models were considered, where one need to
calculate character in presence of only one cycles which is equal to
dimension of the corresponding representation and thus a free fermi
description arises. For a generic matrix model, there exists no
compact expression for character in presence of arbitrary number of
cycles. The closest and most useful formula for character is given by
Frobenius and as we shall see, this formula gives an expression for
character for arbitrary cycles up to integration over some auxiliary
variables.

In this paper we show that although it is difficult to find the most
dominant Young tableaux density for a generic unitary matrix model,
but one can definitely find a relation between number of boxes $h_i$
in a representation and eigenvalues $\q_i$ of a unitary matrix. This
relation defines the boundary of free fermi phase space with constant
density. In the next section we see that this relation allows us to
provide a phase space distribution of unfolded Riemann
zeros\footnote{In a follow up paper we shall present a complete
  construction of Young tableaux distribution for a generic single
  placquette model and show that universality of free fermi
  description for any phases of solution. However, finding the
  boundary relation is enough for our current purpose.}.

Partition function corresponding to single placquette model
(\ref{eq:placquette}) is given by,
\be\label{eq:partfuncplaq}
\cZ = \int [\cD U] \exp\lB N \sum_{n=1}^{\infty}{\beta_{n}\over n}
(\Tr U^{n}+\Tr U^{\dagger n})\rB.
\ee
Expanding the exponential one can write partition function as,
\ben
\cZ = \int [\cD U] \sum_{\vec k} \frac{\varepsilon(\vec \b,\vec k)}
{z_{\vec k}} \Upsilon_{\vec k}(U)
\sum_{\vec l} \frac{\varepsilon(\vec \b,\vec l)}
{z_{\vec l}} \Upsilon_{\vec l}(U^{\dagger})
\een
where,
\ben
\varepsilon(\vec \b, \vec k) = \prod_{n=1}^{\infty}N^{k_n}\b_{n}^{k_n},\quad
z_{\vec k} = \prod_{n=1}^{\infty} k_{n}! n^{k_n}\quad \text{and} \quad 
\Upsilon_{\vec k}(U)=\prod_n (\Tr U^n)^{k_n}.
\een
Here, $n$ runs over positive integers and
$\vec k = (k_1, k_2, \cdots )$ (similarly
$\vec l = (l_1, l_2, \cdots )$), $k_n$ or $l_n$'s can be $0$ or any
positive integer.

$\Upsilon_{\vec{k}}(U)$ can be rewritten in terms of the characters of
the conjugacy class of the permutation group $S_{K}$ as follows
\begin{eqnarray}
\Upsilon_{\vec{k}}(U)=\sum_{R}\chi_{R}(C(\vec{k}))\Tr_{R}[U] .
\end{eqnarray}
Here $\chi_{R}(C(\vec{k}))$ is character of conjugacy class
$C(\vec{k})$ of permutation group $S_{K}$, $K=\sum_\a \a k_\a$ and $R$
denotes representation of $U(N)$. Finally, using following
orthogonality relation between characters of representations of $U(N)$
\begin{eqnarray}
\int {\cal{D}}U\, \Tr_{R}[U] \Tr_{R'}[U^{\dagger}]=\delta_{RR'}
\end{eqnarray}
we obtain the following form for the partition function
\ben
\cZ = \sum_R \sum_{\vec k} \frac{\varepsilon(\vec \b,\vec k)}
{z_{\vec k}} \sum_{\vec l} \frac{\varepsilon(\vec \b,\vec l)}
{z_{\vec l}} \chi_{R}(C(\vec{k}))\chi_{R}(C(\vec{l})).
\een

Any representation of an unitary group $U(N)$ can be labeled by a
Young diagram with maximum $N$ number of rows and arbitrary numbers of
boxes in each row up to a constraint that number of boxes in a
particular row can not be greater than number of boxes in a row before
. Therefore, sum of representations of $U(N)$ can be cast as sum of
different Young diagrams. If $K$ is total number of boxes in a Young
diagram with $\lambda_i$ number of boxes in $i^{th}$ row and
$\sum_i \l_i =K$, then sum over representations can be decomposed as
\begin{eqnarray}
\sum_{R} \rightarrow\sum_{K=1}^{\infty}
\sum_{\{\l_{i}\}}\delta\lb\sum_{i=1}^{N}\l_{i}-K\rb .
\end{eqnarray}
The partition function, therefore, can be written as,
\ben\label{eq:pffinal}
\cZ=\sum_{\vec \l} \sum_{\vec k, \vec l} \frac{\varepsilon(\vec \b, \vec k)
\varepsilon(\vec \b, \vec l)}{z_{\vec k} z_{\vec l}}
\chi_{\vec\l}(C(\vec k)) \chi_{\vec \l}(C(\vec l))
\delta(\Sigma_n n k_n-\Sigma_i \l_i) \delta(\Sigma_n n l_n-\Sigma_i \l_i).
\een
Note that total number of boxes is same as order of permutation group
$S_K$.  The characters of conjugacy class are determined recursively
by Frobenius formula \cite{fulton-harris, hamermesh,lasalle}. Explicit
expressions for the most general case are not simple. We shall discuss
about that in the next subsection.

In large $N$ limit, we introduce continuous variables
\be
h(x) = {h_i\over N}, \quad \text{where}\quad x={i\over N},
\quad \text{with}\quad x\in[0,1]
\ee
where $h_i$ is given by,
\begin{eqnarray}\label{eq:h-nrelation}
h_i = \l_i + N -i \qquad \forall \quad i =1, \cdots, N
\end{eqnarray}
with
\be\label{hmonotonicity}
h_1> h_2> \cdots > h_N \geq 0.
\ee
Therefore, in large $N$ limit partition function can be written
in the following form
\be\label{eq:pfyt1}
\cZ = \int [\cD h(x)] \prod_n\int dk_n' dl_n' \exp\lB -N^2 
S_{\text{eff}} [h(x),\vec {k_n'},\vec{l_n'}]\rB,
\ee
where $S_{\text{eff}}$ is some effective action constructed out of
$h(x)$ and other parameters. Dominant contribution to partition
function comes from those representations which maximise effective
action $S_{\text{eff}}$ and those representations are characterized by
a quantity called Young tableaux density defined as,
\be
u(h) = - {\dow x\over \dow h}.
\ee
Since $h(x)$ is a monotonically decreasing function of $x$, Young
density $u(h)>0 \ \forall \ x\in[0,1]$.

For a simple model, for example, $\b_1\neq 0$ and other $\b_n=0$ (for
$n\geq 2$) it is possible to find $u(h)$ corresponding to different
phases \cite{Dutta:2007ws, dutta-dutta}. In this simple case one
needs to calculate character of permutation group in presence of one
cycles only. Character in presence of non-zero one cycles is given by
dimension of the corresponding representation. However, in presence of other
non-zero cycles it is hard to compute character of permutation
group. In fact there exists no exact expression for that. Therefore,
it is difficult to find $u(h)$ by extremising the effective action for
a generic unitary matrix model.

In that simple model, \cite{Dutta:2007ws, dutta-dutta} observed that
there exists a beautiful identification between Young tableaux
distribution and eigenvalue distribution,
\ben
h \leftrightarrow \r(\q),\quad \text{and} \quad
u(h) \leftrightarrow
\q.
\een
This identification allows one to write eigenvalue distribution and
Young tableaux distribution in terms of a single constant phase space
distribution function $\omega(h,\q)$ such that,
\ben
\r(\q) = \int_0^\infty \o(h,\q) dh, \qquad
u(u) = \int_{-\pi}^\pi \o(h,\q) d\q
\een
where $\o(h ,\q)$ is a distribution in a two dimensional phase space
spanned by $h$ and $\q$ and obeys,
\ben
\begin{split}
\o(h,\q) &= {1\over 2\pi} \ ; \quad (h, \q) \in R\\
&=0 \ ; \quad \text{otherwise.}
\end{split}
\een
It is actually the shape (or boundary) of the region $R$ which
contains all the information about two distributions for a given phase
of the model.  For other phases of solution (for example, one-gap or
two-gap solution), a similar identification holds between eigenvalue
description and Young tableaux description and hence phase space
distribution.

The identification, observed in \cite{Dutta:2007ws, dutta-dutta},
between two distribution was somewhat accidental. There was no clear
understanding behind this.  Specially, it was not known if such
identification holds for a generic model like (\ref{eq:partfuncplaq}).
In this section we show that, although it is difficult to find $u(h)$
for a generic model but identification between $h$ and $\r(\q)$ holds
for this most generic case also.


\subsection{Character of permutation group}\label{character}

We discuss in detail Frobenius character formula
\cite{fulton-harris} for $\chi_{\vec{\l}}(C(\vec k))$ of 
permutation group.  Let $C({\vec{k}})$ denotes the conjugacy class of
$S_K$ which is determined by a collection of sequence of
numbers
\begin{eqnarray}
\vec{k}=(k_{1},k_{2},\cdots), \qquad \text{with}, \ \
\sum_{n}n k_{n}=K .
\end{eqnarray}
$C({\vec{k}})$ consists of permutations having $k_{1}$ number of
1-cycles, $k_{2}$ number of 2-cycles and so on.  We introduce a set of
independent variables, $x_{1},\cdots,x_N$, and for a given Young
diagram $(\vec{\l})$ we have $\lambda_{1}\geq\lambda_{2}\geq \cdots
\geq \lambda_{N}\geq 0$. Then we define a power series and 
Vandermonde determinant as follows.
\begin{eqnarray}
  P_{j}(x)=\sum_{i=1}^{N}x_{i}^{j}\quad\quad
  \quad\Delta(x)=\prod_{i<j}(x_{i}-x_{j}).
\end{eqnarray}
For a set of non-negative integers, $(n_{1},\cdots ,n_{N})$, we define
\begin{equation}
  \lB f(x)\rB_{(n_{1},..,n_{N})}=\textrm{coefficient of }
  x_{1}^{n_{1}}\cdots x_{N}^{n_{N}}\textrm{ in }f(x) .
\end{equation}
%
%
The character corresponding to a conjugacy class $C(\vec k)$ and
a representation characterized by $\vec \l$ is given by famous
Frobenius formula
\begin{eqnarray}\label{eq:Frobenius}
  \chi_{\vec h }(C_{\vec{k}})=\bigg[\Delta(x).\prod_{j} \lb P_{j}(x)
  \rb^{k_j}\bigg]_{(h_{1},\cdots, h_{N})} 
\end{eqnarray}
where $h_i$'s are give by equation (\ref{eq:h-nrelation}).
This is the most generic formula for character. Using this formula one
can in principle write down partition exactly.

In presence of only once cycles the character is nothing but dimension
of the representation $R$. In presence of other cycles, it is difficult to
extract an exact expression for character.


\subsection{Saddle point equations from character}

Writing partition function (equation (\ref{eq:pffinal})) in terms of
representations of $U(N)$ group apparently a different approach in
comparison with writing partition function in terms of eigenvalues of
$U(N)$ matrices (equation \ref{eq:pfeige}). Large $N$ value of
partition function depends of large $N$ behavior of
characters. However, it turns out, with our surprise, that large $N$
behavior of character is controlled by saddle point equation of
eigenvalue distribution. Not only that, upon integrating over
representations one can show that partition function
(\ref{eq:pffinal}), in fact, boils down to equation
(\ref{eq:pfeige2}).

From equation (\ref{eq:pffinal}) it is obvious that for a given
set of $\b_n$s, large $N$ behavior of partition function is controlled
by the same of character. Writing down the Frobenius formula explicitly
we get
\ben
\chi_{\vec h}(C(\vec k)) = \lB \lb\sum_{i=1}^{N}x_{i}\rb^{k_{1}}
\lb\sum_{i=1}^{N} x_{i}^{2}\rb^{k_{2}} \cdots
\prod_{i<j}(x_{i}-x_{j})
\rB_{(h_{1},..h_{N})}.
\een
Introducing $N$ complex variables $(z_1, z_2, \cdots z_N)$, 
character can be written as,
\begin{eqnarray}\label{eq:character2}
 \chi_{\vec h }(C_{\vec{k}})=
\prod_{i}^{N}\oint\frac1{2\pi i}\frac{dz_{i}}{z_{i}^{h_{i}+1}}
\lB \lb\sum_{i=1}^{N}z_{i}\rb^{k_{1}}
\lb\sum_{i=1}^{N} z_{i}^{2}\rb^{k_{2}} \cdots
\prod_{i<j}(z_{i}-z_{j})
\rB.
\end{eqnarray}
The contour is taken around origin. Note that, in the above expressions
for character (equation \ref{eq:character2} or \ref{eq:character2}) 
variables $x_i$'s or $z_i$'s are auxiliary variables. They do not 
carry any physical meaning as of now.

Integrand in the above 
expressions can be exponentiated and written as,
\begin{eqnarray}
\chi_{\vec h }(C_{\vec{k}})=\prod_{i}^{N}\oint
\frac1{2\pi i}\frac{dz_{i}}{z_{i}}
\exp\lB \sum_{n=1}^{\infty}k_{n}\ln(\sum_{i=1}^{N}
z_{i}^{n})+\frac{1}{2}\sum_{i\ne j}\ln|z_{i}-
z_{j}|-\sum_{i=1}^{N}h_{i}\ln z_{i}\rB
\end{eqnarray}
In large $N$ limit we define following variables
\begin{eqnarray}
{z_{i}\over N}= z(x),\quad\quad {h_{i}\over N}=h(x), \quad\quad x={i\over N}
\quad\quad k_{n}=N^{2}k'_n.
\end{eqnarray}
In large $N$ limit summation over $i$ is replaced by an integral over $x$,
\be
\sum_{i=1}^{\infty} \ra N\int_0^1 dx.
\ee
Therefore, in large $N$ limit character is given by,
\ben\label{eq:character1}
\chi[h(x), C(\vec{k'})]= \lb\frac1{2\pi i}\rb^N 
\oint {[\cD z(x)]\over z(x)}
\exp[-N^2 S_{\chi}(h[x],\vec{k'})]
\een
where {\it action} $S_{\chi}$ is given by\footnote{Note that, this is not a 
regular action.} 
\ben\label{eq:Schi}
-S_{\chi}(h[x],\vec{k'}) &=& \sum_n k_n' \lB (n+1)\ln N
+\ln \lb \int_0^1 dx z^n(x)\rb\rB
+\frac12 \int_0^1 dx \Xint -_0^1 dy \ln|z(x)-z(y)|\nonumber\\
&& \quad -K' \ln N
-\int_0^1 dx h(x) \ln z(x)
\een
where, $N^2 K'= K = \text{total number of boxes in a representation}$.

In large $N$ limit,  dominant contribution to character comes
from a particular distribution of $z(x)$, which extremises the function
$S_{\chi}$. Extremisation condition is given by,
\ben
\sum_n {n k_n'\over Z_n} z^n(x) + \Xint-_0^1 dy
{z(x)\over z(x)-z(y)}
-h(x) =0
\een
where,
\be
Z_n = \int_0^1 dx z^n(x).
\ee
Introducing,
\be
\r(z)={\partial x\over \partial z}
\ee
above equation can be written as,
\ben\label{eq:saddlez}
\sum_n {n k_n'\over Z_n} z^n + \Xint- dz'
{z \r(z') \over z-z'}
-h(z) =0.
\een
%
%
%
%
%
Since original contour in equation (\ref{eq:character2}) was
about the origin, therefore we take $z=e^{i\theta}$ and above equation
can be written as,
\be\label{eq:chaeqn}
\sum_n {n k_n'\over Z_n} e^{i n \theta} 
+\Xint-_{-\pi}^{\pi} d\theta' \r(\theta') \frac{e^{i\theta}}
{e^{i\theta}-e^{i\theta'} }  -h(\q)=0
\ee
where,
\be
\r(\theta) = {\dow x\over \dow \q} = i e^{i\q} {\dow x\over \dow z}
= i e^{i\q} \r(z).
\ee
Breaking this equation into real and imaginary part we find
that the imaginary part is given by,
\ben
\sum_n {n k_n'\over Z_n}\sin{n\q} 
-\frac12 \Xint-_{-\pi}^{\pi} d\theta' \r(\theta') \cota{{\q-\q'\over 2}}=0.
\een
This equation is exactly same as eigenvalue equation
(\ref{eq:eveqnplaq}) once we identify, $n k'_n/Z_n = \b_n$.
In the next subsection we shall see that the above identification 
is correct and auxiliary
variables which appear in Frobenius formula are indeed eigenvalues
of unitary matrices under consideration. These auxiliary variables
satisfy not only the same equation as eigenvalues, partition
function written in terms of these variables exactly matches with
partition function written in terms of eigenvalues. Hence, in  
large $N$ limit values of $k_n'$ (number of cycles) are fixed
in terms of parameters of the model {\it i.e.} $\vec \b$ and  
density $\r(\q)$ defined above is same as the density of 
eigenvalues.

The real part of equation (\ref{eq:chaeqn}) is an algebraic equation,
which relates number of boxes in a particular Young diagram with 
eigenvalues of unitary matrices,
\be
h(\q)=\frac12 + \sum_n {n k_n'\over Z_n}\cos{n\q} .
\ee

In terms of $\theta$ and $\r(\q)$ action $S_\chi$ (\ref{eq:Schi}) 
is given by,
\ben\label{eq:Schi2}
\begin{split}
-S_{\chi}(h[x],\vec{k'}) &= \sum_n k_n' \lB (n+1)\ln N
+\ln \lb \int_{-\pi}^{\pi} d\q \r(\q) \cos n\q\rb\rB \\
&+\frac12 \int_{-\pi}^{\pi} d\q \r(\q)\Xint -_{-\pi}^{\pi} d\q'
\r(\q')\ln|e^{1\q}-e^{i\q'}|
-K' \ln N
-i \int_{-\pi}^{\pi} d\q \r(\q)h(\q)\q
\end{split}
\een
Since $\r(\q)$ is even function of $\q$, the last terms implies that
there exists a redundancy in $h(\q)$. For any even function $f(\q)$
$$S_\chi\lB h(\q),\vec k\rB = S_\chi\lB h(\q)+f(\q),\vec k\rB.$$ This
means that different Young distributions which are related by this
transformation correspond to the same eigenvalue distribution
$\r(\q)$. Hence, the most generic relation between $h(\q)$ and $\q$ is
given by,
\be\label{eq:generichtheta}
h(\q) = \frac12 + \sum_n {n k_n'\over Z_n}\cos{n\q}+f(\q)
\ee
in large $N$ limit. This redundancy plays an important role to give a
phase space description of different phases of a unitary matrix model.
We name this equation "boundary equation" as it defines the boundary
of phase space.  We shall get back to this point in section
\ref{sec:redun}. Before that we explicitly show auxiliary variables
$\q_i$'s are indeed eigenvalues of unitary matrices. They satisfy not
only the same saddle point equation but partition function written in
terms of these auxiliary variables is same as the partition function
written in terms of eigenvalues.


\subsection{The partition function  at large N}

Using the expression for character given in equation (\ref{eq:character1})
partition function (\ref{eq:pfyt1}) can be written as,
\ben\label{eq:pftotal}
\cZ = \cN \prod_n \int dk_n' dl_n' \int \cD h(x)\oint \cD z(x)
\oint{\cal{D}}w(x) \int_{\I}^\I dt \int_\I^\I ds
\exp\lb - N^2 S_{\text{total}}\lB h,z,w,\vec{k'},
\vec{l'},t,s\rB \rb\nonumber\\
\een
where,
\begin{eqnarray}\label{eq:effStotal}
\begin{split}
- S_{\text{total}}\lB h,z,w,\vec{k'},
\vec{l'},t,s\rB &= \sum_n \bigg[ k_n'\lb1+ \lna{{\b_n Z_n\over n k_n'}}\rb 
+l_n'\lb1+ \lna{{\b_n W_n\over n l_n'}}\rb \bigg]\\
&\qquad +\frac12 \int_0^1 dx \Xint-_0^1 dy\lb \ln|z(x)-z(y)|
+\ln|w(x)-w(y)|\rb\\
& \qquad -\int_0^1 dx h(x) \ln z(x) w(x) 
+i t \lb \sum_n n k_n' -K'\rb +i s \lb \sum_n n l_n' -K'\rb.
\end{split}
\end{eqnarray}
Varying effective action with respect to $h(x)$ we find,
\be
z(x)w(x) = e^{-i(t+s)} = \text{constant (independent of} \ x).
\ee
Variation with respect to $k_n'$ and $l_n'$ provides,
\ben
{\b_n Z_n\over n k_n'} = e^{-i t n}, \quad \text{and}
\quad {\b_n Z_n \over n l_n'} = e^{-i s n}.
\een
Since, in equation (\ref{eq:pftotal}) $z(x)$ and $w(x)$ vary over some
contour around zero, we take these contours to be unit circle about
the origin and hence one set of solutions to above equations are,
\be
z(x) = e^{i \q(x)}, \quad w(x) = e^{-i\q(x)}, \quad \text{and} \quad
t=s=0.
\ee
Therefore,
\be
k_n' = l_n'= {\b_n \r_n\over n}, \quad \text{where} \quad
Z_n = \int d\q \r(\q) \cos n\q \equiv\r_n.
\ee
Evaluating the partition function (\ref{eq:pftotal}) on these 
equations we find,
\begin{eqnarray}\label{eq:effStotal}
\begin{split}
\cZ &= \cN \int [\cD \q] \exp\lB N^{2}\sum_{n=1}^{\infty}
2\frac{\b_n}{n}\r_n
+N^{2}\frac{1}{2}\Xint-d\q\rho(\q)\Xint-d\q'
\rho(\q')\ln\left| 4\sin^{2}\lb\frac{\q-\q'}{2}\rb\right|\rB.
\end{split}
\end{eqnarray}
This partition function is exactly same as the partition function of
one placquette model (equation \ref{eq:pfeige2}) written as an
integral over eigenvalues of unitary matrices. Hence, we identify that
the auxiliary variables we used to write the character of symmetric
group as eigenvalues of corresponding unitary matrix model involved.


\subsection{Redundancy in boundary equation and phase space
  distribution}
\label{sec:redun}

It is evident either from equation (\ref{eq:Schi}) or
(\ref{eq:effStotal}) that in large $N$ limit there is a redundancy in
Young tableaux distribution. However, there is further restriction on
this redundancy, which comes from the fact that redundancy can not
change the topology of phase space distribution given by a constant density. In fact, this
restriction on redundancy actually fixes phase space distribution. Let
us consider tow cases.

\noindent {\bf No-gap phase:}\ For no-gap, phase space distribution is
given by a region around the origin, i.e. for a given $\q$ $h$ has two
solutions $0$ and $h_+(\q)=2\pi\r(\q)$.  From equation
(\ref{eq:generichtheta}) we get,
\be
h(h-S(\q))=f^2(\q)-\frac14 S^2(\q),
\ee
where,
\be
S(\q) = 1 +2\sum_n \frac{nk_n'}{Z_n}\cos n\q.
\ee
Therefore, $h$ has solution $0$ and $\r(\q)$ provided
\be
S(\q) = 2\pi \r(\q)\qquad \text{and} \qquad 
f(\q) = \frac12S(\q)=\pi\r(\q).
\ee

\noindent {\bf Multi-gap phase :}\ Similarly for multi-gap solution,
topology of phase space distribution implies that there exists two
non-zero solution for $h$, denoted by $h_+$ and $h_-$ and
\be\label{eq:identification-1cut}
h_+(\q)-h_-(\q) = 2\pi \r(\q).
\ee
Againg, from equation (\ref{eq:generichtheta}) we get,
\be
h^2 - S(\q)h + \frac14 S^2(\q)-f^2(\q)=0
\ee
which has solution
\be\label{eq:hpm}
h_{\pm}(\q)=\frac12S(\q)\pm f(\q).
\ee
Hence, equation (\ref{eq:identification-1cut}) implies,
\be
f(\q)=\pi \r(\q).
\ee

One can, in fact, construct Young distribution for each phase
considering the identification $\q=\pi u(h)$ from this boundary
relation. However, the complete construction is not necessary in our
current analysis and hence we shall present that in details in a
follow up paper.


\subsection{A generalised phasespace distribution}



It had been observed in \cite{Dutta:2007ws, dutta-dutta} that for a
particular class of unitary model (namely $(a,b)$ model), eigenvalue
distribution and Young tableaux distribution can be obtained from a
single distribution function. This function, which takes values either
zero or one in the entire two dimensional plane ($(h,\q)$ plane), was
identified with the phase space distribution of free fermions in one
dimension. In this section we show that it is indeed possible to get
all the information about eigenvalue and Young tableaux distribution
of different phases of a unitary matrix model of type
(\ref{eq:partfuncplaq}) from a single phase space distribution.

We define a complex phasespace distribution $\O(h,\q)$ in $(h,\q)$
plane such that
\be\label{eq:gen-phasespace}
\int_0^\infty \O(h,\q) dh = 2R(e^{i\q})-1.
\ee
Being complex one can break the function $\O(h,\q)$ into real and
imaginary part
\be
\O(h,\q) = \o_\Re (h,\q) + i \o_\Im(h,q).
\ee
When integrated over $h$, real part gives eigenvalue distribution
where as imaginary part gives derivative of potential.
\be
\int_0^\infty dh \ \o_\Re(h,\q) = \r(\q),\quad 
\int_0^\infty dh \ \o_\Im(h,\q) =  V'(\q).
\ee
Since, $\r(\q)$ and $V'(\q)$ are even and odd function of $\q$
respectively, the real and imaginary part of $\O(h,\q)$ are also real
and imaginary respectively. Hence, when we integrate $\O(h,\q)$ over
$\q$ only the real part contributes and gives Young tableaux
distribution function
\be
u(h) = \int_{-\pi}^{\pi} \Omega(h,\q) d\q.
\ee

As mentioned in the beginning of this section, free fermi phases
spaces are two dimensional droplets in $(h,\q)$ plane with constant
density inside each droplets. Therefore, real part of $\Omega(h,\q)$
is given by,
\be
\omega_{\Re} (h,\q) = \frac1{2\pi} \Theta[(h-h_-(\q))(h_+(\q)-h)]
\ee
where $h_{\pm}(\q)$ is given by equation (\ref{eq:hpm}).

Imaginary part of $\Omega(h,\q)$ can also be written in terms of
constant density droplets in $(h,\q)$ plane
\ben
\begin{split}
\omega_{\Im}(h,\q) &=\ \ \ \Theta(h-\sum_n \b_n \sin n\q), \quad 0\leq \q
\leq \pi \\
&= - \Theta(h+\sum_n \b_n \sin n\q) , \quad -\pi \leq \q
\leq 0.
\end{split}
\een
Note that the shape or topology of imaginary part of $\Omega(h,\q)$ is
fixed for a given model. This is because the imaginary part contains
information about derivative of potential, which is fixed for a
particular model. Whereas, topology of real part of $\Omega(h,\q)$
changes as we go from one phase to another phase.

%% file: phspace-prime.tex
\section{Phase space distribution of Riemann zeros}
\label{sec:phasespace}

\subsection{Multiple Young distribution corresponding to Riemann zeros
}

\begin{figure}[h]
\begin{subfigure}{.4\textwidth}
\centering
\includegraphics[width=4cm,height=3cm]{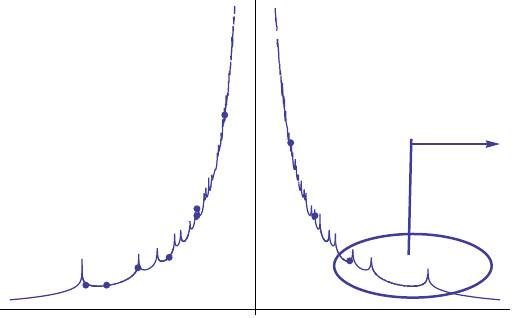}
\end{subfigure}
\begin{subfigure}{.5\textwidth}
\centering
\includegraphics[width=9cm,height=5cm]{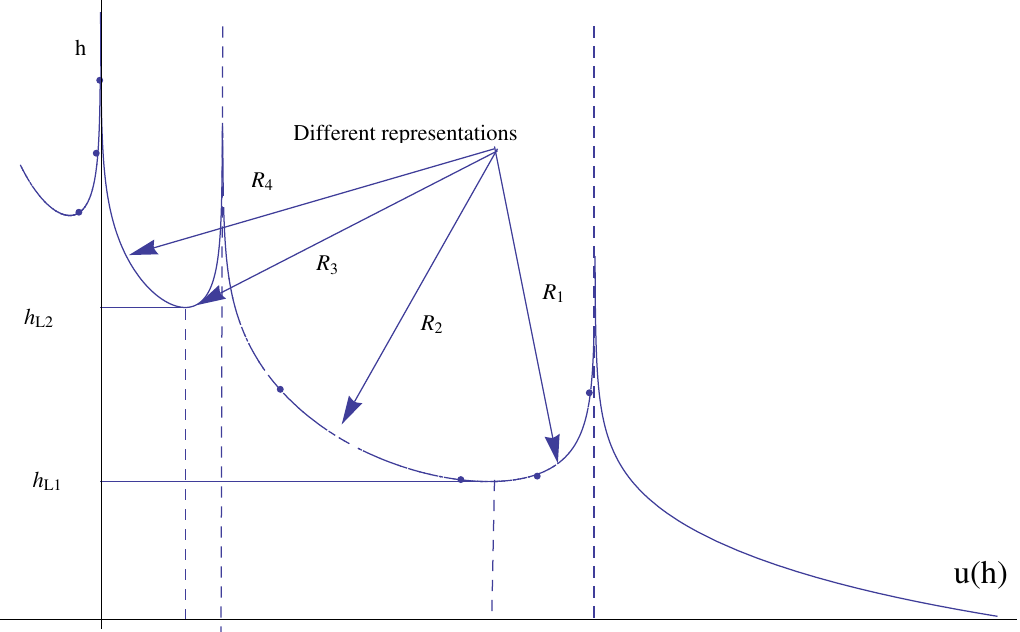}
\end{subfigure}
\caption{Young tableaux distribution for Riemann zeros. We see that 
there are multiple representations for a given $h$.}
\label{fig:multiuh}
\end{figure}

In this section we shall find Young distribution for no-gap solution
which captures information about distribution of unfolded zeros of
\rzf, discussed in section \ref{sec:model}. Since eigenvalue
distribution, given by equation (\ref{eq:evdensity}), has no gap and
hence belongs to no-gap phase. Therefore, boundary relation for this
phase is given by,
\ben
\bsp
h &= \frac12 +\sum_n \b_n \cos n \q +\pi \r(\q)\\
&= \frac12 -\a\sum_n {\l_n\over n} \cos n \q +\pi \r(\q)\\
&=2\pi \r(\q).
\end{split}
\een
From the plot of $\r(\q)$ (fig. \ref{fig:rhoplot}) it is clear that
the relation is not one two one. For a given $h$ there are multiple
possible values of $\q$. Since $\q$ is identified with density of
Young distribution $\pi u(h)$, this implies there exists many possible
Young representations which correspond to the same eigenvalue
distribution. Each of these representations are defined such that $h$
varies monotonically from a lowest value to a highest value. For
example, consider a particular sector of the plot \ref{fig:rhoplot} as
shown in figure \ref{fig:multiuh}.
Representation $R_1$ is defined as,
\ben
\bsp
u_{R_1}(h) &= \a_1, \quad h\in[0, h_{L1}]\\
&= u(h), \quad h\in[h_{L1},\infty].
\end{split}
\een
Similarly representation $R_2$ is defined as,
\ben
\bsp
u_{R_2}(h) &= \a_2, \quad h\in[0, h_{L1}]\\
&= u(h), \quad h\in[h_{L1},\infty]
\end{split}
\een
and so on. The constants $\a_1, \ \a_2$ etc. are fixed from the
normalisation condition 
\be
\int_0^{\infty}u(h)\ dh =1
\ee
for each representation.

Thus we see that, for a given eigenvalue distribution there exists
multiple representations of $U(N)$ group which extremises the
character and the partition function.

\subsection{Phase space distribution of Riemann zeros}

%
%
%
%
%
%
%
%
%
%
%
%
%
%

\begin{figure}[h]
\centering
\includegraphics[width=10cm,height=7cm]{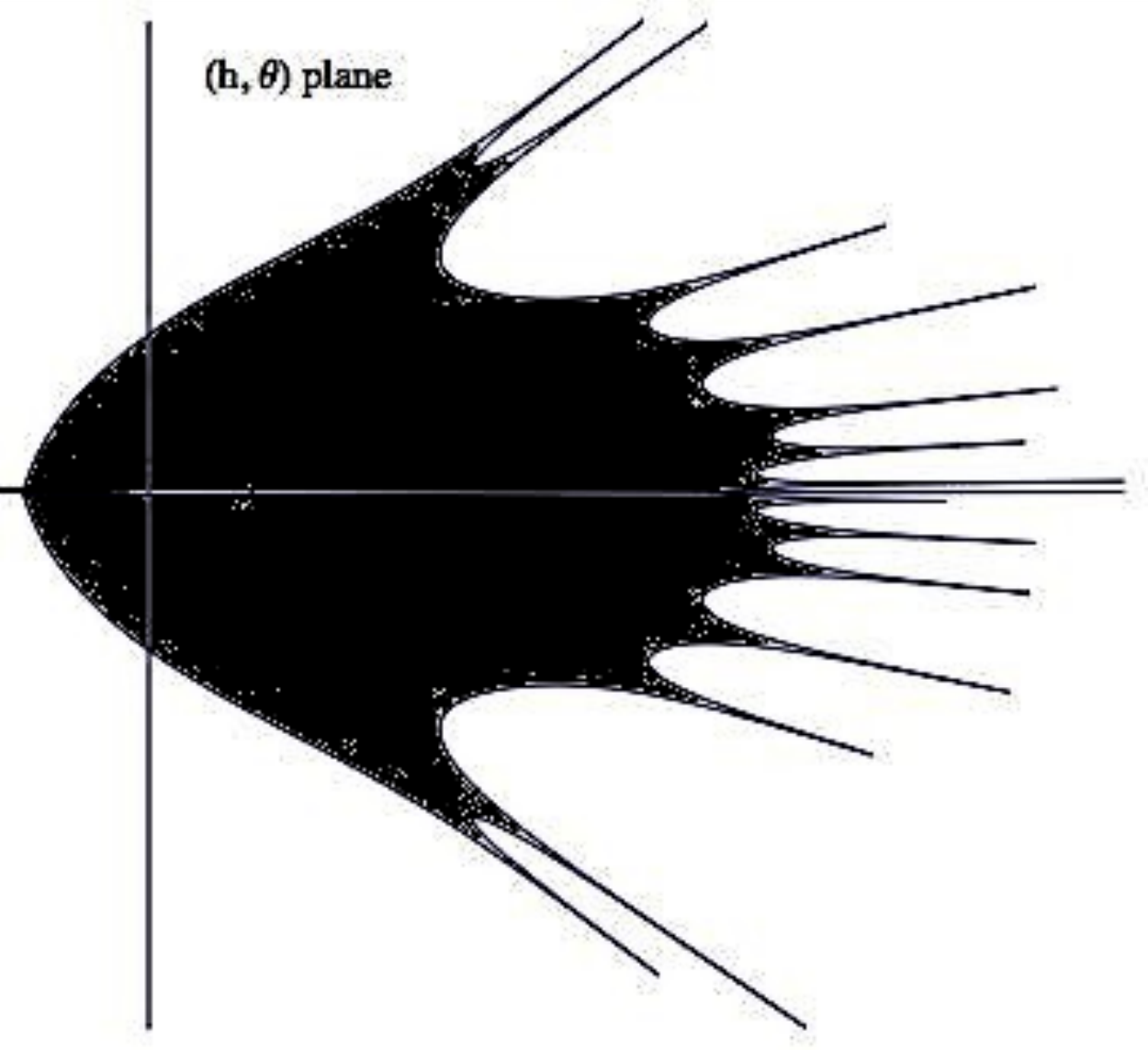}
\caption{Phase space distribution of Riemann zeros.}
\label{fig:phasespace}
\end{figure}

Since boundary of phase space for no-gap solution is given by
$h=\r(\q)$ one can draw a phase space distribution in $(h,\q)$
plane. Since $h>0$ we take $h$ as radial variable and $\q$ periodic.
The distribution is plotted in fig. (\ref{fig:phasespace}).

We see that the phase space looks like broom with sticks at different
positions of Riemann zeros.

\subsection{Young distributions and prime counting function}

Resolvent $R(z)$ is written in terms of generalised distribution
\be\label{eq:R-Omega}
R(e^{i\q})= \frac12 +\int_0^\infty \Omega(h, \q) \ dh.
\ee
To understand a relation between Young distribution and distribution
between prime numbers we introduce a function $J(x)$, as defined in
\cite{hmedwards}, which captures information about distribution of
prime numbers.
\begin{eqnarray}
J(x)={\rm Li}(x)-\sum_{i}{\rm Li}(x^{\rho_{i}})-  \ln
  2+\int_{x}^{\infty} \frac{dt}{t(t^{2}-1)}\quad\quad x>1
\end{eqnarray}
where, ${\rm Li}(x)$ is logarithmic integral function defined as,
\be
{\rm Li}(x) =\int_0^{x}\frac{dt}{\ln t}
\ee
and $\r_i$'s are zeros of \rzf. Function $J(x)$ is a monotonically
increasing function and increases by a jump of $1/n$ at each $p^{n}$ where
$p$ is a prime number.

Riemann showed that zeta function $\zeta(s)$ and prime counting
function $J(x)$ are related to each other by,
\ben
\begin{split}
\frac{\ln \z(s)}{s} &= \int_0^\infty J(x) x^{-s-1}dx \qquad \qquad
\Re[s]>1\\
J(x)&=\frac1{2\pi i } \int_{a-i \infty}^{a+\infty} \ln \z(s) x^s
\frac{ds}{s} \qquad a>1.
\end{split}
\een
Using this relation one can write,
\begin{eqnarray}\label{eq:xi-pc}
\ln\xi\lb\frac{1}{2}+\frac{i}{2}\cot\frac{\theta}{2}\rb=\int_{0}^{\infty}
  \,x^{-\frac{1}{2}-\frac{i}{2}\cot\frac{\theta}{2}}\frac{d\tilde{J}}{dx}dx
\end{eqnarray}
where $\tilde{J}(x)$ is fluctuating part of $J(x)$ and explicitly
given in terms of zeroes of zeta function \cite{hmedwards}
\be 
d\tilde{J}(x)=-\frac{1}{\ln
  x}\sum_{\rho_{i}}x^{\rho_{i}-1}dx\notag.
 \ee
Hence, from equation (\ref{eq:xi-pc}) we find that,
\begin{eqnarray}
\begin{split}
  -2\alpha\ln\xi\lb\frac{1}{2}+\frac{i}{2}\cot\frac{\theta}{2}\rb&= 2R(e^{i\q})-1\\
  &= -2\a \int_{0}^{\infty} \,[x(\tilde
  J)]^{-\frac{1}{2}-\frac{i}{2}\cot\frac{\theta}{2}}d\tilde{J}\\
  &= -2\a \int_0^\infty \tilde \Omega(\tilde J, \q) \ d\tilde J,
  \qquad \text{where} \quad \Omega(\tilde J, \q)=[x(\tilde
  J)]^{-\frac{1}{2}-\frac{i}{2}\cot\frac{\theta}{2}}.
\end{split}
\end{eqnarray}
where the limit in the integral is kept over $x$ for simplicity. Comparing this result with equation (\ref{eq:R-Omega} and
\ref{eq:gen-phasespace}) we see a surprising similarity between the
variables $\tilde J$ and $h$. Eigenvalue density can be obtained from
a distribution function $\tilde \Omega(\tilde J, \q)$ defined on
$(\tilde J,\q)$ plane and integrating over $\tilde J$. The difference
is, in $(h,\q)$ plane it the boundary of droplets which determines the
eigenvalue density whereas in $(\tilde J, \q)$ plane distribution
function $\tilde \Omega(\tilde J,\q)$ determines $\r(\q)$.

In $(\tilde J,\q)$ plane one can also integrate over $\q$ and find,
\begin{eqnarray}
\begin{split}
  \frac{1}{2\pi}\int_{-\pi}^{\pi}d\q \ \tilde \Omega(\tilde J,\q)&=
  \frac{1}{2\pi}\int_{-\pi}^{\pi}d\theta [x(\tilde
  J)]^{-\frac{1}{2}-\frac{i}{2}\cot\frac{\theta}{2}}\\
&=[x(\tilde J)]^{-1}.
\end{split}
\end{eqnarray}
Since $\tilde J(x)$ is fluctuation part of $J(x)$, mapping between $x$
and $\tilde J$ is not one to one. For a given values of $\tilde J$
there are infinite solution for $x$. Therefore, we can split $x$ into
domains where the map ${\tilde{J}} \ra x$ is one to one which is
analogous to what happened for the case of Young Tableaux distribution
function from the boundary equation. Also, since $1<x<\infty$ we have
$1 > [x({\tilde J})]^{-1} > 0$. Thus we see an interesting similarity
between Young distribution for \rzf\ and distribution of prime numbers
\be
h\leftrightarrow \tilde J, \qquad u(h) \leftrightarrow 
[x({\tilde J})]^{-1}.
\ee

It is interesting to see that for this particular choice of the phase
space density we have $\tilde{J}$ playing the role of $h$ and
$x^{-1}\sim\theta$. This suggest that there exists a mapping between these
variables which bear resemblance to canonical transformations in phase
space descriptions. Notably since $\tilde{J}$ is related to the
Fourier transform of $\ln\xi$ and Fourier transform preserves the
symplectic structure of Hamiltonian mechanics, this further provides
evidence of an underlying phase space description of the nontrivial
zeroes of the zeta function and the prime numbers. We intend to look
into this problem in our future work.

%% file: appendix.tex
\appendix

\section{Calculation of $\rho(\theta)$ }\label{app:evcalcu1}

\ben
\rho(\theta) &=& -2\a \Re \lB \ln \xi ({1\over 1-e^{i\theta}})\rB\nonumber\\
&=& -\a \ln\lB \xi({1\over 1-e^{i\theta}}) \xi({1\over 1-e^{-i\theta}})\rB
\nonumber\\
&=& 2\a\ln 2 -\a \sum_i \ln \lB {\r_i(1-e^{i\theta})-1 \over \r_i(1-e^{i\theta})}
{\r_i(1-e^{-i\theta})-1 \over \r_i(1-e^{-i\theta})  }\rB\nonumber\\
&=& 2\alpha\ln2 -\a \sum_i \ln \lB {1-e^{i(\theta+\theta_i)}\over 1- e^{i\theta}}\bigg]\bigg[
{1-e^{i(\theta_i-\theta)}\over 1- e^{-i\theta}}\rB
\een
Replacing $\theta_{i}$ by $-\theta_{i}$ and simplifying we obtain:
\begin{eqnarray}
  \rho(\theta)&=&1-\alpha\sum_{i}\ln\bigg[e^{i\frac{(\theta-\theta_{i})}{2}}
                  \frac{(e^{-i\frac{(\theta-\theta_{i})}{2}}-e^{i\frac{(\theta-\theta_{i})}{2}})}
                  {e^{i\frac{\theta}{2}}(e^{-i\frac{\theta}{2}}-e^{i\frac{\theta}{2}})}\bigg]
                  \bigg[e^{-i\frac{(\theta-\theta_{i})}{2}}\frac{(e^{-i\frac{(\theta-\theta_{i})}{2}}-
                  e^{i\frac{(\theta-\theta_{i})}{2}})}{e^{-i\frac{\theta}{2}}(e^{-i\frac{\theta}{2}}
                  -e^{i\frac{\theta}{2}})}\bigg]\notag\\
              &=&2\alpha\ln2-\alpha\sum_{i}\ln\bigg[\frac{\sin^{2}(\frac{\theta
                  -\theta_{i}}{2})}{\sin^{2}\frac{\theta}{2}}\bigg]
\end{eqnarray}
putting the value of $\alpha=\frac{1}{2\ln2}$ we obtain the result.

\section{Normalisation condition for $\r(\theta)$}
\label{app:rhonormalization}

The following integral can also be checked in Mathematica :

\ben
\int_{-\pi}^{\pi} \r(\theta) d\theta &=& \int_{-\pi}^{\pi}
\lB 1+2\a \sum_i \ln \sin\lb {\theta \over 2}\rb 
-2\a \sum_i \ln \sin \lb {\theta-\theta_i \over 2}\rb\rB d\theta
\nonumber\\
&=& 2\pi + 2\alpha \sum_i \lb i \pi^2 -2\pi \ln 2\rb
-2\a \sum_i \lb i \pi \theta_i +i \pi^2 -2\pi \ln 2\rb\nonumber\\
&=& 2\pi -2\pi i \a \sum_i \theta_i\nonumber\\
&=& 2\pi.\nonumber
\een
Where we have used the fact that the $\theta_{i}$ appear symmetrically
about the real line and hence for every $\theta_{i}$ there is a
$\theta_{i}$ and thus their sum is $0$. Equivalently this can be
checked using the normalization of $R(z)$ :
\begin{eqnarray}
\frac{1}{2\pi i}\oint\frac{dz}{z}R(z)=1.
\end{eqnarray}

\section{Identification $\theta=\pi u(h)$}
\label{app:iden}
Since the auxiliary variables in the Frobenius character formula were identically mapped to the eigen values of the Unitary matrix models we have the condition : 
\begin{eqnarray}
\int dx\, \theta(x)h(x)=0
\end{eqnarray}
This means $h(x)$ is always an even function of $\theta(x)$. Now we start with the action defined in \ref{eq:Schi2} and add an additional term 
\begin{eqnarray}\label{multip1}
\int dx\,\mu(x)\frac{\partial x}{\partial h(x)}\h(x)
\end{eqnarray}
where $\mu$ can be understood as a Lagrange multiplier which enforces the constraint $u(h)>0$. Collecting the $h(x)$ dependent terms only, we have the following:
\begin{eqnarray}
\int dx\, \theta(x)h(x)+\int dx\,\mu(x)\frac{\partial x}{\partial h(x)}\theta(x)
\end{eqnarray}
Since $h(x)$ is always positive we can write $h=\gamma^{*}\gamma$ following \cite{Jurkiewicz:1982iz}. Variation with respect to the phase part of $\gamma^{*}$ yields the equation of motion:
\begin{eqnarray}
&&\theta(x)+\mu(x)\frac{\partial x}{\partial h(x)}=0\notag\\
&&-\frac{\partial x}{\partial h(x)}=\frac{1}{\mu(x)}\theta(x)
\end{eqnarray}
Inserting this in \ref{multip1} we always have the value of the integral $0$ as $h(x)$ has to be an even function of $\theta(x)$. Now for positivity of $u(h)=-\partial x/\partial h$ we must have $1/\mu(x)$ to be an odd function of $\theta(x)$ and for $u(h)<1$ we must have $|\mu|>\pi$. For the particular choice $\mu=\pi{\rm sign}(\theta)$ we have $u(h)=\frac{1}{\pi}\theta$, which is the identification obtained. Thus the multiplier $\mu$ shows us that the identification $\pi u(h)=\theta$ satisfies all the constraints needed. 

%
%